\documentclass[aps,prb,twocolumn,superscriptaddress,floatfix,showkeys,longbibliography]{revtex4-2}
\usepackage{etoolbox}
\usepackage[english]{babel}
\usepackage{graphics,graphicx} 
\usepackage{natbib}
\usepackage[labelformat=parens]{subfig}
\usepackage{float}
\usepackage{placeins} 
\usepackage{ragged2e} 
\usepackage{caption}
\usepackage{xcolor}
\usepackage{amsmath}
\usepackage{amssymb}
\usepackage{braket}
\usepackage{blindtext}
\usepackage{bm}
\usepackage{algorithm}
\usepackage{algpseudocode}
\usepackage{hyperref}
\hypersetup{
    colorlinks=true,
    citecolor=blue,
    linkcolor= red,
    urlcolor= blue 
}
\usepackage[labelformat=parens]{subfig}
\usepackage[justification=raggedright, font=footnotesize]{caption}

\usepackage{MnSymbol}
\usepackage{xurl}

\usepackage[cal=boondox,scr=boondoxo]{mathalfa}

\begin{document}

\title{Many-body neural network wavefunction for a non-Hermitian Ising chain}

\author{Lavoisier Wah}
\email[]{lavoisier.wahkenounouh@mpl.mpg.de}
\affiliation{Max Planck Institute for the Science of Light, 91058 Erlangen, Germany}
\affiliation{Department of Physics, Friedrich-Alexander-Universit\"at Erlangen-N\"urnberg, 91058 Erlangen, Germany}

\author{Remmy Zen}
\email[]{remmy.zen@mpl.mpg.de}
\affiliation{Max Planck Institute for the Science of Light, 91058 Erlangen, Germany}

\author{Flore K. Kunst}
\email[]{flore.kunst@mpl.mpg.de}
\affiliation{Max Planck Institute for the Science of Light, 91058 Erlangen, Germany}
\affiliation{Department of Physics, Friedrich-Alexander-Universit\"at Erlangen-N\"urnberg, 91058 Erlangen, Germany}

\date{\today}

\begin{abstract}
Non-Hermitian (NH) quantum systems have emerged as a powerful framework for describing open quantum systems, non-equilibrium dynamics, and engineered quantum optical materials. However, solving the ground-state properties of NH systems is challenging due to the exponential scaling of the Hilbert space, and exotic phenomena such as the emergence of exceptional points. Another challenge arises from the limitations of traditional methods like exact diagonalization (ED). For the past decade, neural networks (NNs) have shown promise in approximating many-body wavefunctions, yet their application to NH systems remains largely unexplored. In this paper, we explore different NN architectures to investigate the ground-state properties of a parity-time-symmetric, one-dimensional NH, transverse field Ising model with a complex spectrum by employing a recurrent neural network~(RNN), a restricted Boltzmann machine~(RBM), and a multilayer perceptron~(MLP). We construct the NN-based many-body wavefunctions and validate our approach by recovering the ground-state properties of the model for small system sizes, finding excellent agreement with ED. Furthermore, for larger system sizes, we demonstrate that the RNN outperforms both the RBM and MLP. However, we show that the accuracy of the RBM and MLP can be significantly improved through transfer learning, allowing them to perform comparably to the RNN for larger system sizes. These results highlight the potential of neural network-based approaches—particularly for accurately capturing the low-energy physics of NH quantum systems in case of both weak and strong non-Hermiticity.

\end{abstract}

\maketitle

\section{Introduction}\label{S0}
Traditional quantum mechanics relies on Hermitian Hamiltonians to ensure real eigenvalues and unitary time evolution. However, non-Hermitian (NH) quantum mechanics has emerged as a powerful framework to describe open quantum systems, where interactions with the environment may lead to energy dissipation and gain \cite{ashida2020non,bergholtz2021exceptional}. A particularly intriguing subclass of non-Hermitian systems consists of those obeying parity-time ($PT$) symmetry, where the Hamiltonian remains invariant under simultaneous spatial reflection and time-reversal operations \cite{bender1998real}. Such systems exhibit real eigenvalues in the unbroken $PT$-symmetric phase, and undergo a phase transition through an exceptional point (EP), where two or more eigenvalues and their corresponding eigenvectors coalesce. Beyond this point, the system enters the spontaneously broken $PT$-symmetric regime, where eigenvalues become complex \cite{el2018non}. This rich phenomenology has inspired studies across condensed matter physics, quantum optics, and cold-atom systems \cite{ateuafack2025probing, zhang2025non, bergholtz2021exceptional, Ozdemir2019, el2018non, Li2019, Ren2022, Ding2021}.

Recent advances have extended NH physics from single-particle systems to many-body quantum systems, unveiling exotic phenomena and a richer understanding of EPs and the NH skin effect leading to different phases~\cite{Gohsrich2025}, unconventional localization and distinct quantum criticality \cite{Luitz2019, Zhang2020, lu2024many,  Sun2021}. These developments have been explored in various contexts, including NH extensions of the transverse field Ising model (TFIM), where complex interactions lead to unconventional symmetry breaking and dynamical properties \cite{lu2024many, lenke2021high, li2014conventional, guerra2025correlations}. Despite significant progress in theoretical formulations, an efficient numerical method to determine the ground-state properties of non-Hermitian many-body systems remains an open challenge due to the exponential scaling of the Hilbert space as well as the instabilities induced by the emergence of EPs.

The application of artificial intelligence and machine learning techniques in quantum many-body physics has provided innovative approaches to studying strongly correlated systems \cite{krenn2023artificial,carleo2019machine, carrasquilla2020machine}. Neural-network-based variational wavefunctions or neural quantum states (NQSs), such as restricted Boltzmann machines (RBMs), multilayer perceptrons (MLPs), and recurrent neural networks (RNNs) have demonstrated remarkable success in approximating ground states of complex quantum systems~\cite{carleo2017solving,hibat2020recurrent, pilati2019self, bayram2014study,pilati2019self, 10.21468/SciPostPhys.10.6.147,westerhout2020generalization, schmitt2025simulating,rende2025foundation}. By parametrizing the wavefunction with a neural network model, it can capture entanglement and correlation structures efficiently, making them well suited for studying quantum phases and transitions \cite{carleo2017solving,hibat2020recurrent,sharir2020deep}. However, NQSs have so far been restricted to Hermitian many-body systems, and their extension to non-Hermitian many-body physics offers a promising alternative to traditional numerical techniques such as exact diagonalization (ED), which become computationally prohibitive for large system sizes. Tensor network methods, while effective in one dimension and capable of treating large systems, are limited in the set of states they can efficiently capture, as compared to NQSs~\cite{sinibaldi2025non}. Moreover, their adaptation to NH settings — especially those involving complex-valued wavefunctions and biorthogonal structures — remains challenging. In contrast, NQSs offer a flexible and expressive representation that can naturally accommodate these complexities and hold significant potential for future extensions to higher-dimensional or strongly entangled systems. Very recently, NQS techniques have only been applied to study NH fermionic systems with real spectrum using the RNN \cite{ibarra2025autoregressive}. Here, we propose to further the study to the realm of many-body NH spin chain with a \emph{complex spectrum}, which is the ``raison d’\^etre'' of this paper. Our goal is to develop fast and efficient algorithms capable of capturing the many-body wavefunction of NH systems  and propose a method to investigate their ground state.

Extending many-body neural network wavefunctions to NH systems enables the exploration of exotic ground states, particularly in $PT$-symmetric models, where standard Hermitian variational techniques may fail. This work aims to develop and apply a neural-network-based variational Monte Carlo~(VMC) method to study the ground-state properties of a $PT$-symmetric non-Hermitian TFIM with complex spectrum. The TFIM is the perfect model to test VMC in NH systems because the model can be solved analytically~\cite{lenke2021high, li2014conventional, STARKOV2023169268}. Using our numerical techniques, we not only find that our results are in perfect agreement with ED but also provide insights into metastable states that are not fully accessible with ED, as discussed later.

The rest of the work is organized as follows. In Sec.~\ref{S1}, we present our model, discuss the concept of unbroken-to-broken phase transition, demonstrate the emergence of EPs, and elaborate on the concept of the ground state in NH systems. In Sec.~\ref{S2}, we demonstrate how to recover the ground state for our model. We then show in Sec.~\ref{S3} how transfer learning (TL) can enhance the performance of the neural networks. Finally, our results are summarized in Sec.~\ref{S5}.

\section{Model and non-Hermitian symmetry}\label{S1}
\subsection{Hamiltonian}

\begin{figure}
    \includegraphics[width=0.48\textwidth]{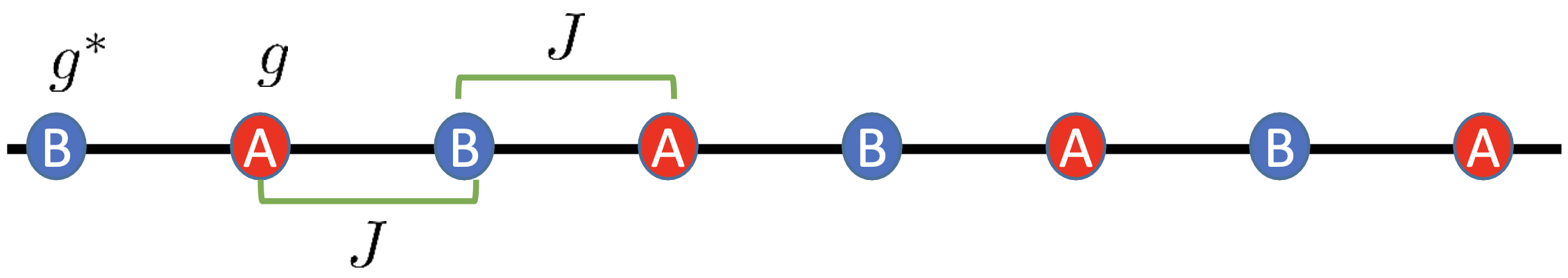}
    \caption{\textbf{1D Ising chain in a staggered magnetic field}. $A$ (red) and $B$ (blue) represent the sublattices in which spin-$1/2$ particles occupy the odd and even positions, respectively. The complex field $g$ ($g^*$) is applied to the $A$ ($B$) sublattices. Neighboring spins are coupled via the coupling constant $J$.   }
    \label{fig1}
\end{figure}

\begin{figure}
    \includegraphics[width=0.40\textwidth]{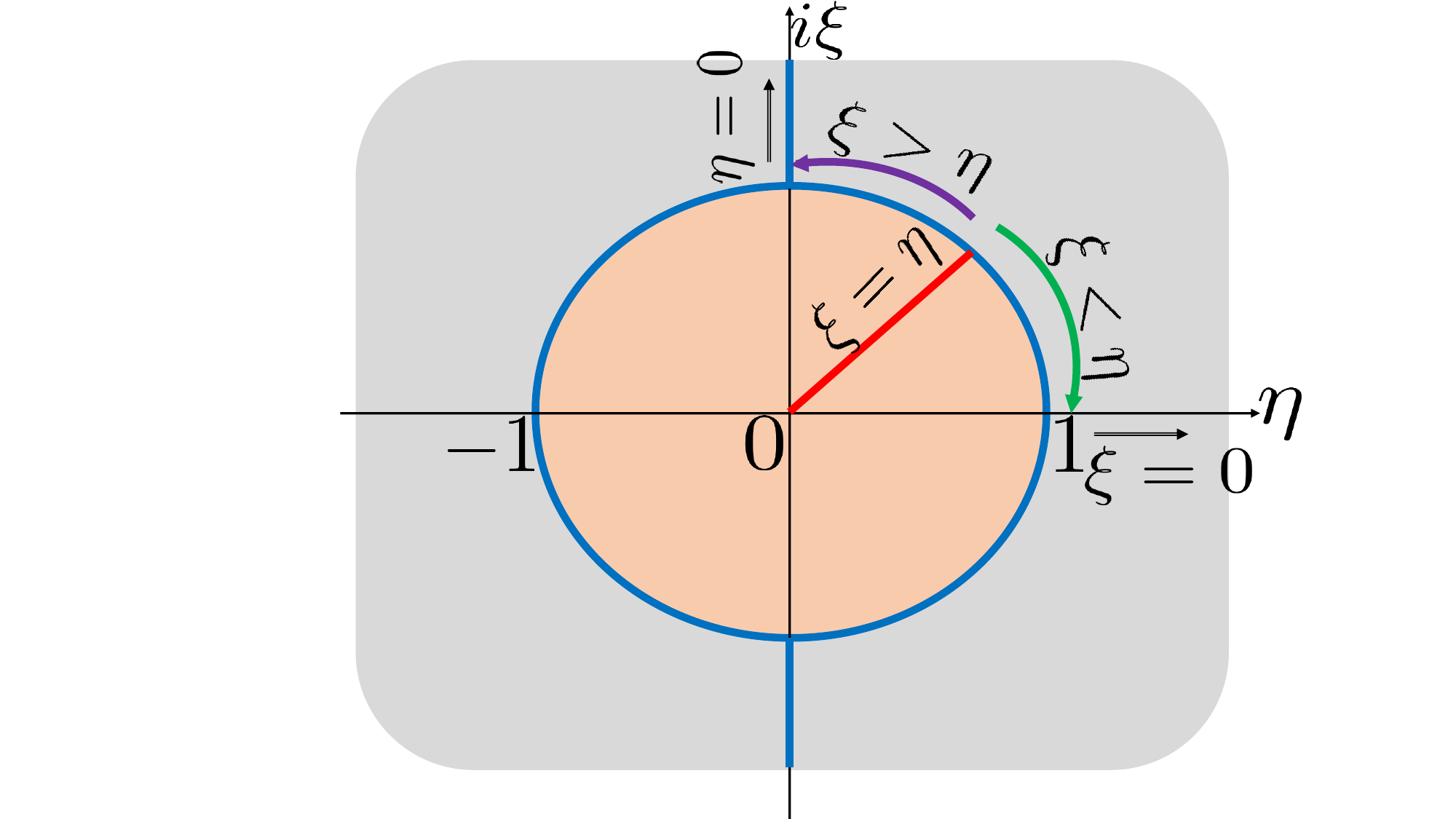}
    \caption{\textbf{Phase diagram of the ground state of the NH TFIM}. The orange region is the ferromagnetic phase, while the gray region represents the paramagnetic phase with the blue circle being the set of critical points at which the quantum phase transition occurs. The plot represents different regimes of the system namely, the regime of weak non-Hermiticity $\xi < \eta$ (green arrow), strong non-Hermiticity $\xi > \eta$ (violet arrow), complete non-Hermiticity  $ \eta=0$, equal contribution of the real and imaginary parts of $g$ (red line), and complete Hermiticity $\xi =0$. }
    \label{fig2}
\end{figure}

\begin{figure*}[!ht]
    \centering
    
    \subfloat[\centering ]{{\includegraphics[width=0.33\textwidth]{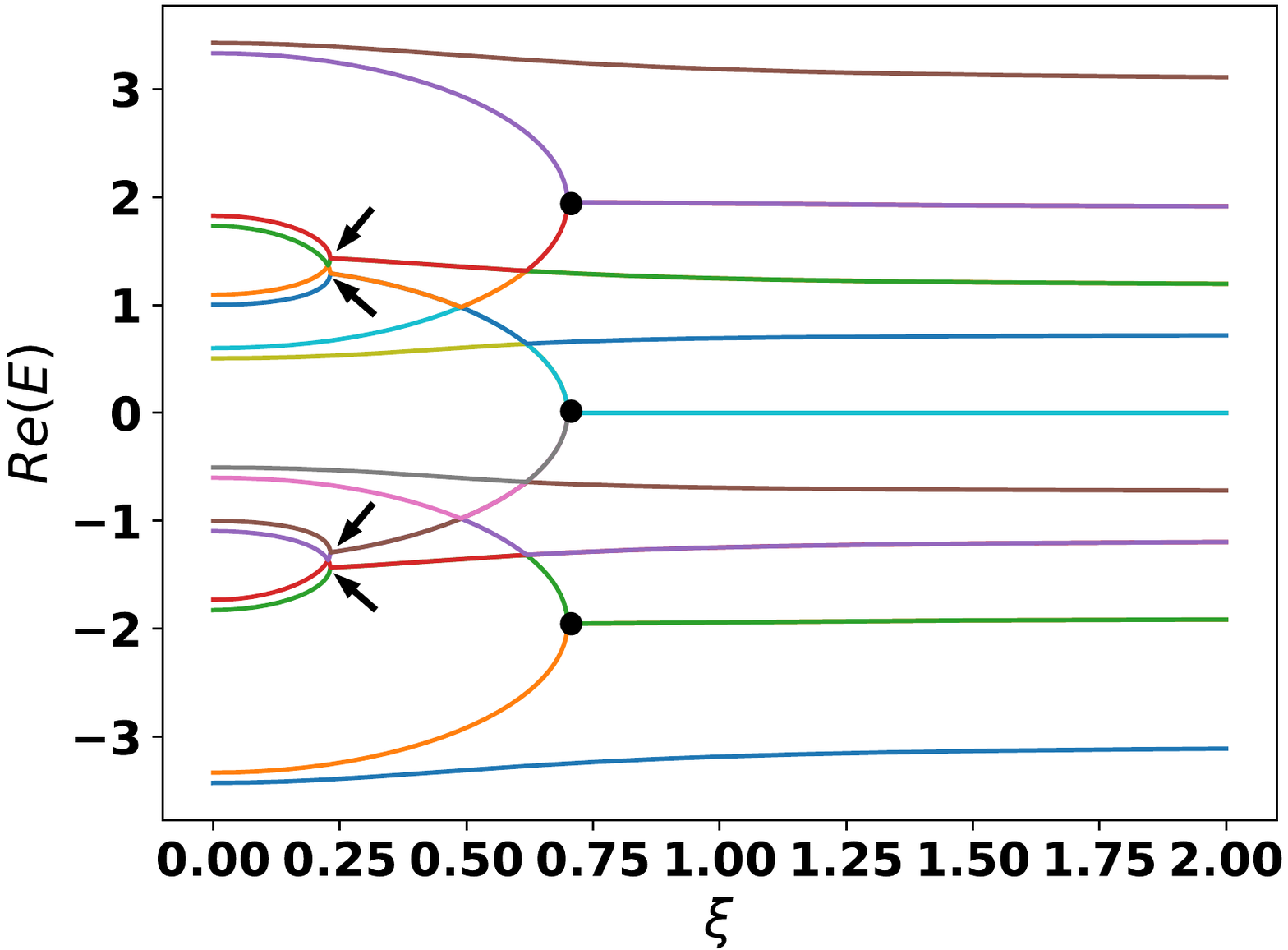} }}
    \subfloat[\centering ]{{\includegraphics[width=0.33\textwidth]{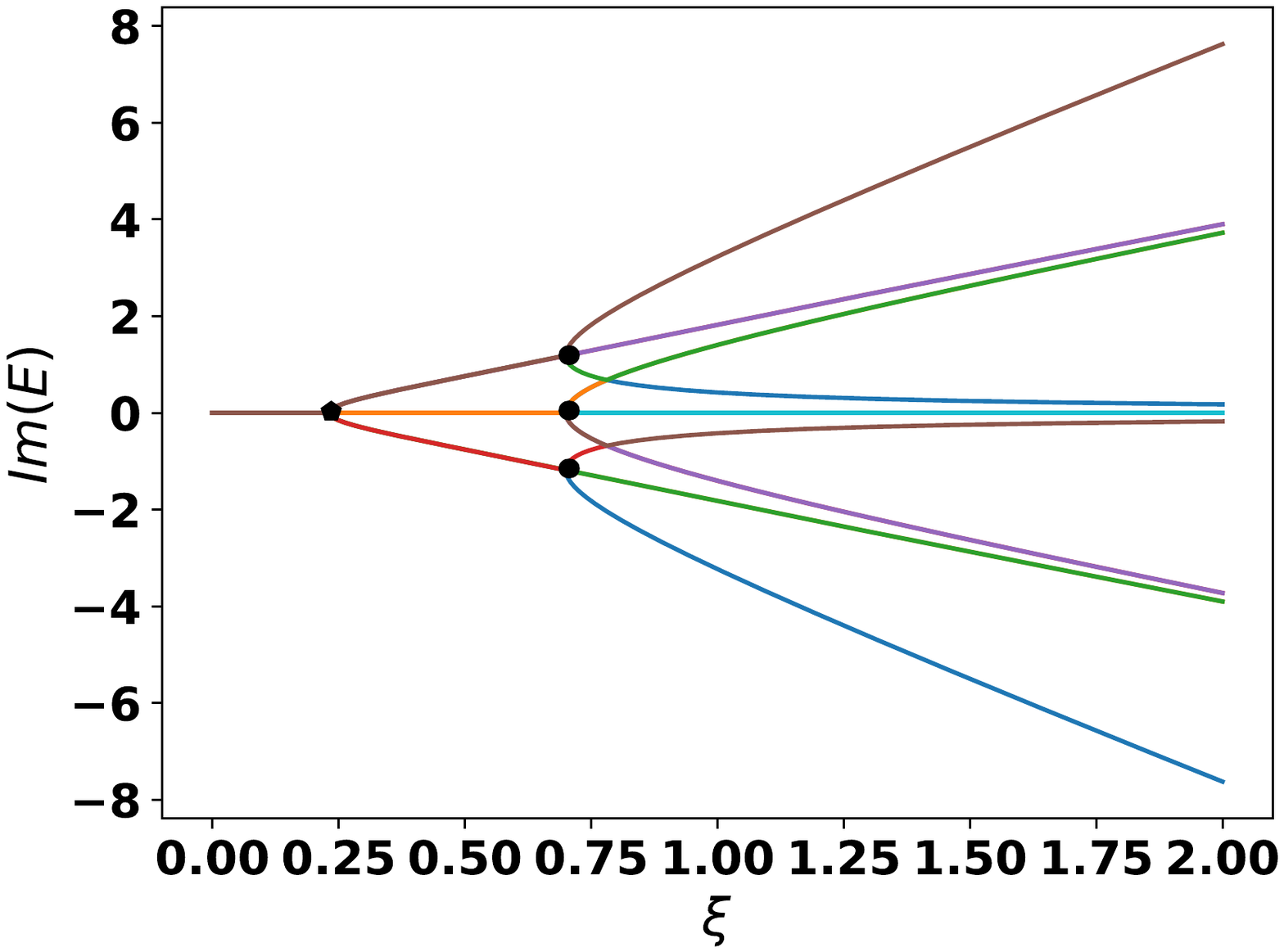} }}
    \subfloat[\centering ]{{\includegraphics[width=0.33\textwidth]{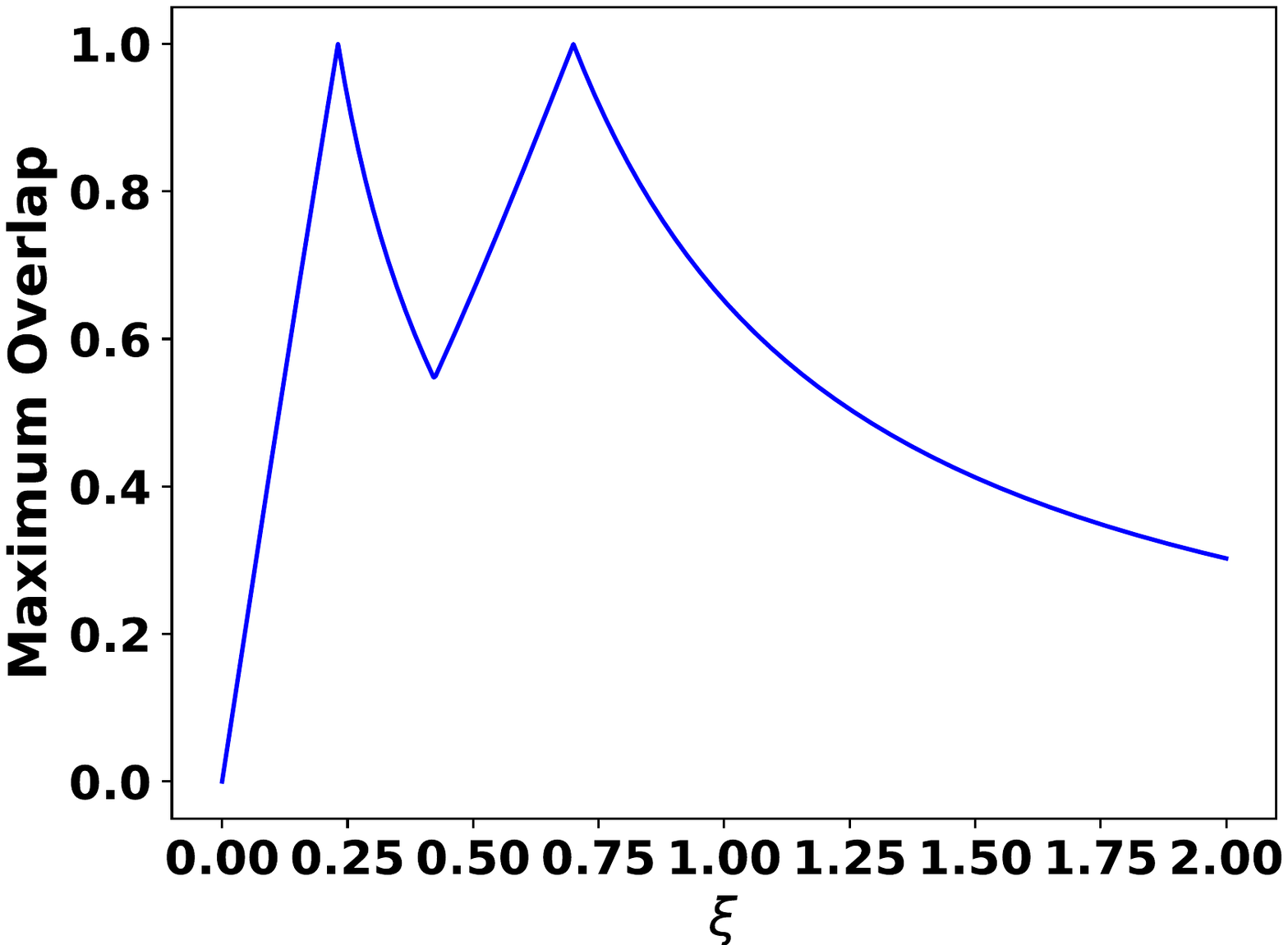} }}
    \qquad
    \subfloat[\centering ]{{\includegraphics[width=0.33\textwidth]{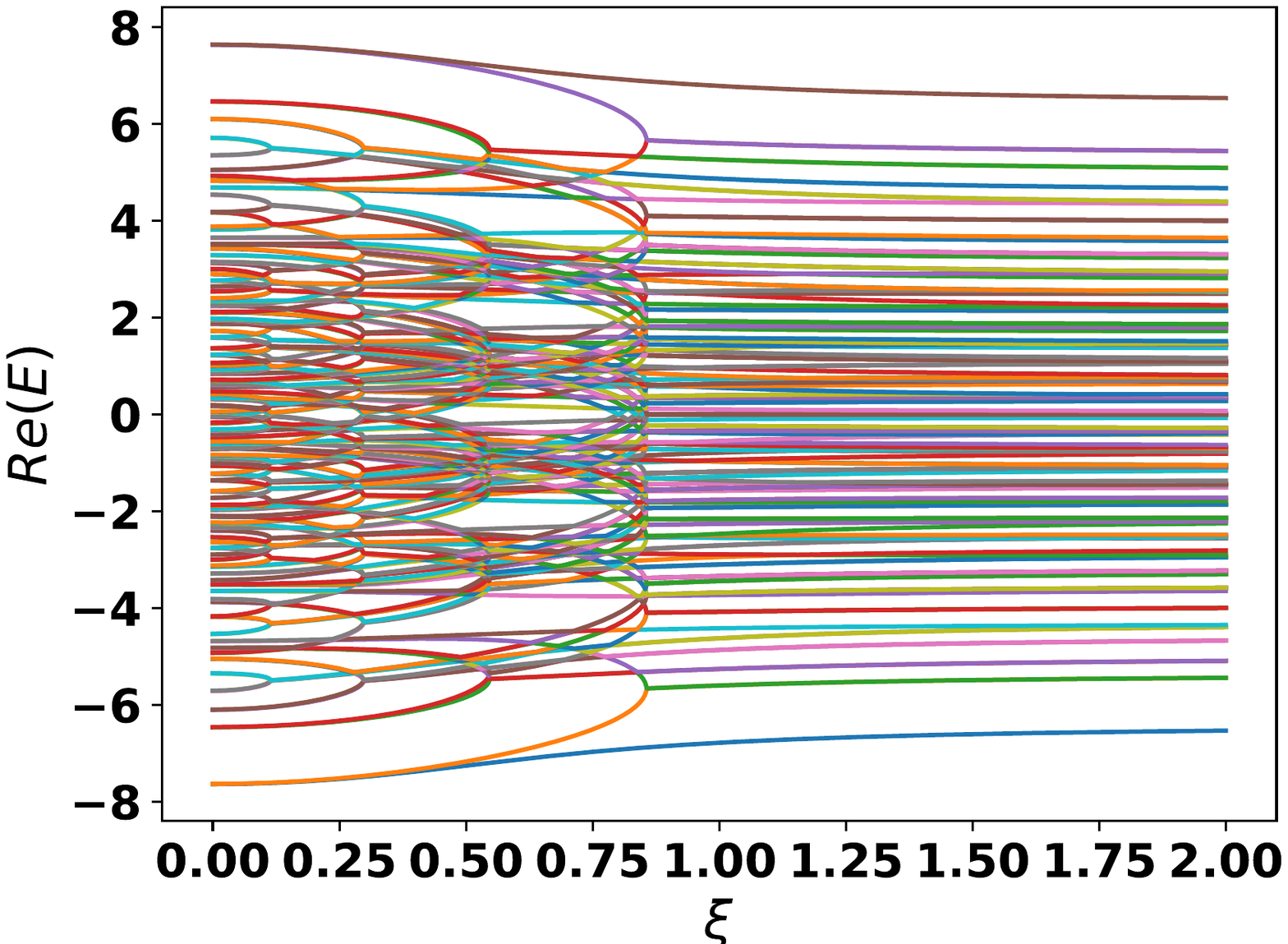} }}
    \subfloat[\centering ]{{\includegraphics[width=0.33\textwidth]{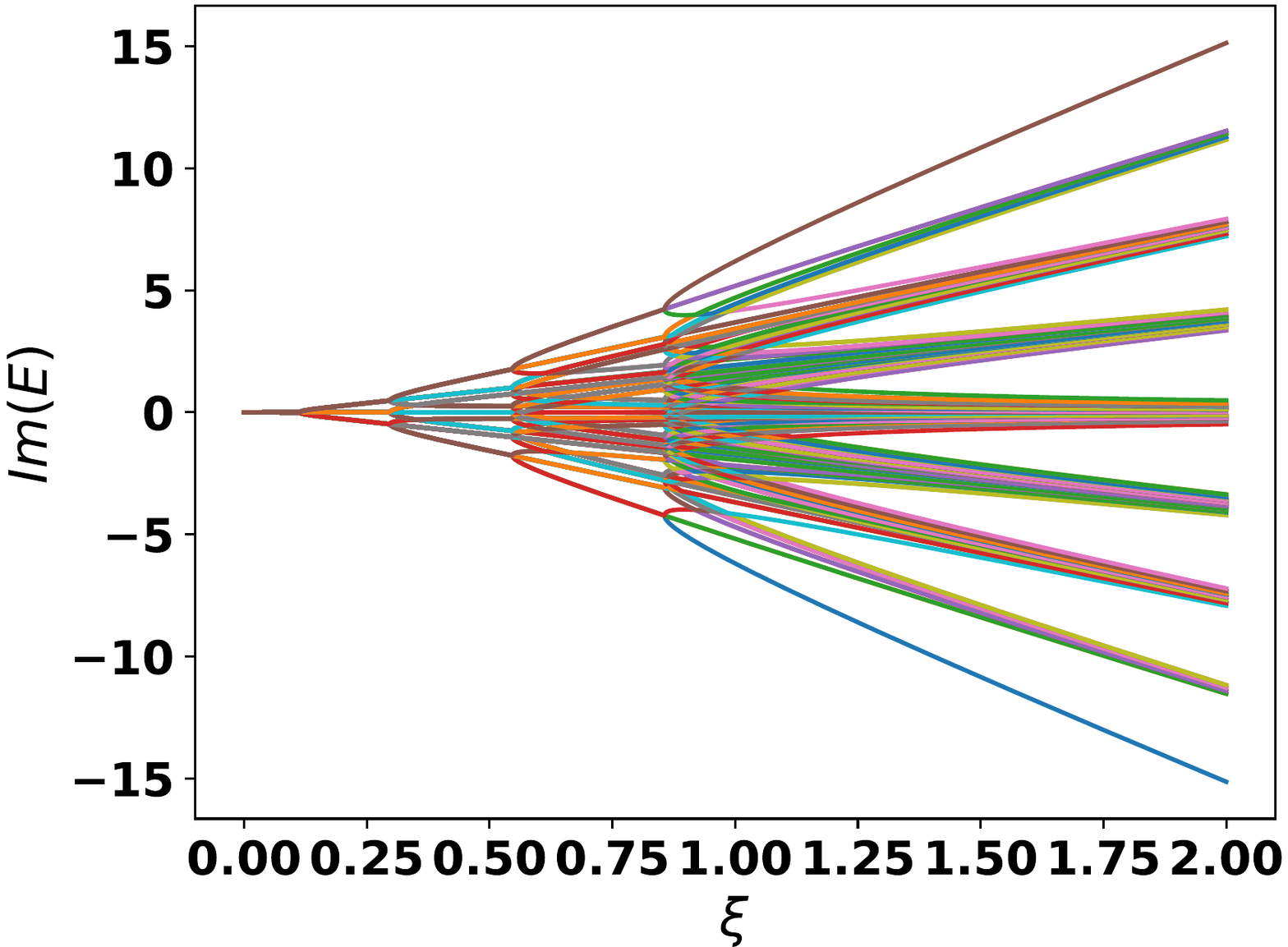} }}
    \subfloat[\centering ]{{\includegraphics[width=0.33\textwidth]{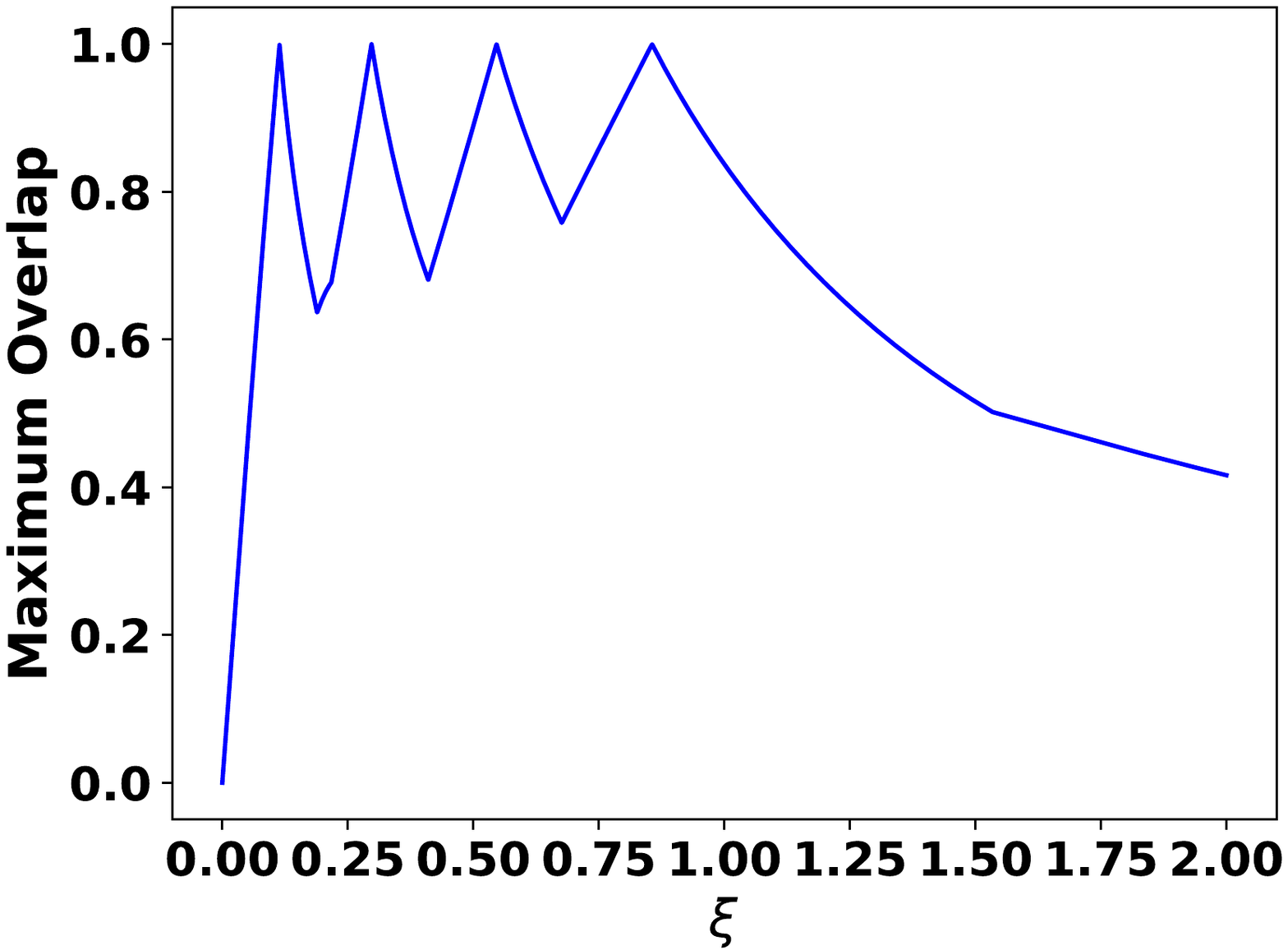} }}
    
    \caption{\textbf{Unbroken-to-broken phase transition: Exceptional points}. We analyze the \textbf{(a)} real and \textbf{(b)} imaginary parts of the energy spectrum of the system and observe the emergence of exceptional points [highlighted with black dots and arrows in panel \textbf{(a)}]. We consider $N=4$ and $\eta=0.5$ and the coloring is determined by the order in which the bands appear. To effectively confirm that these points are EPs, we analyze the \textbf{(c)} maximum overlap between all eigenstates and observe two peaks corresponding to the location of EPs. Likewise for panels \textbf{(d-f)} with $N=8$, one observes the emergence of EPs at four different values of $\xi$. OBCs are considered in this figure. }
    \label{fig9}
\end{figure*}

The model of interest consists of a one-dimensional (1D) quantum Ising chain of $N$ spin-$1/2$ particles on two sublattices $A$ and $B$ coupled via a spin coupling $J$ and subject to a complex, staggered magnetic field that introduces non-Hermiticity into the system (see Fig.~\ref{fig1}). The NH field is chosen so that $g= \eta + i \xi$ is applied perpendicularly to the spins on sublattice $A$ (red) and $g^*= \eta - i \xi$ to the spins on sublattice $B$ (blue) with $\eta, \xi \in \mathbb{R}$. The Hamiltonian describing such a system is given by 
\begin{align}\label{Eq1}
    H = -J \sum_{j=1}^N \sigma_j^z \sigma_{j+1}^z -g \sum_{j \in A}\sigma_j^x-g^* \sum_{j \in B}\sigma_j^x,
\end{align}
where $\sigma_j^z$ and $\sigma_j^x$ are Pauli matrices. This setting imposes that any gain due to the imaginary part of the magnetic field, $+\xi$, is balanced by an equivalent loss, $-\xi$. Moreover, the real part of the field, $\eta$, behaves like a standard transverse field, driving fluctuations between spin states. This model is $PT$-symmetric. The action of the parity operator $P$, which amounts to spacial reflection, is written as $P \sigma_j^{\nu} P^{-1} = \sigma_{N-j+1}^{\nu}$. The time-reversal operator $T$ acts as $T i T^{-1} = -i$, and $T\sigma_j^{\nu} T^{-1} = \sigma^{\nu}_j$ with $\nu =x,z$, that is, it leaves the spin on each site invariant. One can then show that $[PT, H] = 0$. Unless stated otherwise, we shall work in the regime where $J=1$ and periodic boundary conditions (PBCs) are considered, that is, $\sigma_j^\nu = \sigma_{j + N}^\nu$. Under these conditions, the eigenstates of the system immediately break the $PT$ symmetry, while the Hamiltonian remains globally $PT$-symmetric [see the Supplemental Material for details in Ref.~\cite{supmat}]. This phenomenon is known as spontaneous $PT$ symmetry breaking \cite{yu2021spontaneous}. The phase diagram of the system in the complex plane is obtained by plotting the energy density as a function of the components of the field. A projection of this phase diagram in the $(\eta-i\xi)$ plane is depicted in Fig.~\ref{fig2}. As discussed in Ref.~\cite{li2014conventional}, the non-Hermiticity has the effect of shrinking the ferromagnetic phase (orange region). Indeed, the presence of non-Hermiticity introduces a perturbation in the system. In the paramagnetic phase (gray region), the spins are disordered, however, the magnetic field tends to align all spins in the same direction, creating an ordered state. Non-Hermiticity disrupts this alignment, leading to more disordered or unstable behavior. This disruption promotes instabilities by suppressing the stabilizing quantum fluctuations induced by the field resulting in shrinking of the ferromagnetic phase.

It is instructive to note that the blue lines along the $\xi$-axis separating paramagnetic phases on both sides, do not indicate a phase transition but rather a crossover, reflecting a dramatic change in the structure of the system’s ground state~\cite{li2014conventional}. Also, the NH Hamiltonian in Eq.~\eqref{Eq1} can be solved analytically in an analogous fashion to the Hermitian case. We refer to Refs.~\cite{lenke2021high, li2014conventional} for detailed calculations.

\subsection{Unbroken-to-broken phase transition}
In NH systems, particularly those with $PT$ symmetry, the concept of phase transitions is richer and can be extended beyond what is typically encountered in Hermitian systems. Rather than focusing on phase transitions associated with spin alignment, as is generally the case in Hermitian spin models (and which we will discuss later), we begin here by examining a different type of transition: A sudden change in the system's eigenstates. This type of phase transition is called the  \textit{unbroken-to-broken} phase transition, which directly relates to the $PT$ symmetry in the model.

The \textit{unbroken-to-broken} phase transition describes a change between two regimes distinguished by the behavior of the eigenvalues and, crucially, the eigenvectors of the system’s Hamiltonian. 

In the unbroken $PT$-symmetric phase, the Hamiltonian has entirely real eigenvalues [see Fig.~\ref{fig9}(a)], and its eigenvectors are invariant under the $PT$ operation. The reality of the eigenvalues is analogous to traditional Hermitian systems, where all eigenstates correspond to well-defined physical states. In this regime, the system exhibits real and stable quantum evolution (unitary), with physical observables such as magnetization and energy behaving similarly to those in conventional Hermitian systems. Changing the system's parameters may lead to a spontaneous breakdown of $PT$ symmetry enabling the system to enter the broken phase. In this regime, the eigenvalues of the system's Hamiltonian become complex or purely imaginary [see Fig.~\ref{fig9}(b)]~\cite{bender1998real}, the eigenvectors are no longer invariant under $PT$ symmetry, and the system undergoes non-unitary evolution. The transition from real to complex eigenvalues marks the crossing from the unbroken to the spontaneously broken phase and this crossing passes through an EP. The spontaneously broken $PT$-symmetric phase is more akin to a dissipative or unstable regime, where the system may exhibit some exotic behaviors including the emergence of additional EPs (responsible of numerical instabilities), and physical observables can show non-conservative, NH behavior with exponential growth or decay.

From Figs.~\ref{fig9}(a) and ~\ref{fig9}(b), we observe that at specific points, pairs of eigenstates merge with a characteristic square-root behavior indicating the presence of an EP. At these EPs, the two eigenvalues become degenerate and the corresponding eigenvectors also coalesce, which is a defining feature of NH degeneracies.
Immediately beyond this coalescence, the energy spectrum transitions from being entirely real to having complex eigenvalues, which is the hallmark of spontaneous $PT$-symmetry breaking. This transition signifies that the system has entered the spontaneously broken $PT$ phase.

\subsection{Exceptional points}

Under OBCs, we investigate the emergence of EPs arising during the \textit{unbroken-to-broken} phase transition.
The presence of these EPs is first hinted at by the square-root singularity, which is a universal feature of NH phase transitions and is linked to the criticality of the system at the transition point. It can be further confirmed by analyzing the overlap between eigenstates. Indeed, we compute an ``all-to-all'' dot product between all pairs eigenstates of the system  $\mathbf{v}_i$ and $\mathbf{v}_j$ (where $i \neq j$) and calculate their overlaps. The maximum overlap is the largest value among all these pairwise overlaps:
\begin{align}\label{Eq5}
    \text{max\_overlap} = \max_{i \neq j} \left| \mathbf{v}_i^\dagger \mathbf{v}_j \right|.
\end{align}
At an EP, two or more eigenvectors become linearly dependent, meaning they ``merge" or coalesce. As a result, the overlap between these eigenvectors reaches exactly one at the EP. Away from an EP, the eigenvectors remain linearly independent, and the maximum overlap is strictly less than 1, as shown in Fig.~\ref{fig9}(c). This behavior can be summarized as follows:
\begin{align}\label{Eq6}
    \text{max\_overlap} = 1 \quad &\text{(at an EP)}, \notag \\
    \text{max\_overlap} < 1 \quad &\text{(away from an EP)}.
\end{align}  
From Fig.~\ref{fig9}(c), we observe two distinct peaks where the maximum overlap reaches 1. These peaks appear precisely at the locations of the eigenvalue degeneracies, thus confirming that they are indeed EPs. The first peak of the maximum overlap corresponds to the position of four superimposed EPs, while the second peak marks the location of the remaining three EPs. The same observation can be made for a system with $N=8$, cf. Figs.~\ref{fig9}(d-f).

\subsection{Ground state of non-Hermitian systems}
Likewise to a conventional (Hermitian) system, the ground-state (indexed with 0) average of any physical observable in NH systems can be obtained by ``sandwiching'' the Hamiltonian with some eigenstates. In this particular case, one of the most used formulations is the biorthogonal formalism \cite{kunst2018biorthogonal, brody2013biorthogonal}, which can be understood as an NH analog to the quantum mixed estimator in Hermitian physics. As such, for a given observable $\mathcal{O}$ one has
\begin{equation}\label{Eq.2}
    \langle \mathcal{O} \rangle = \frac{\langle \Psi_{0,L}|\mathcal{O}|\Psi_{0,R}\rangle}{\langle \Psi_{0,L}| \Psi_{0,R}\rangle},
\end{equation}
where $|\Psi_{0,R}\rangle$ and $\langle\Psi_{0,L}|$ are the right and left ground-state wavefunctions of $H$, respectively, satisfying the biorthonormal product
\begin{equation}\label{Eq.3}
    \langle \Psi_{i,L}| \Psi_{j,R}\rangle =\delta_{i,j}.
\end{equation}
Following this prescription, for a given spin configuration $\bm{x} =\{x_1, x_2, ..., x_N\}$, we compute the local energy of the system as follows:
\begin{equation}\label{Eq4}
    E_\textrm{loc}(\bm{x})=\frac{\langle\bm{x}|H|\Psi_{0,R}\rangle}{\langle \bm{x}| \Psi_{0,R}\rangle} = \frac{\langle \Psi_{0,L}|H| \bm{x}\rangle}{\langle \Psi_{0,L} | \bm{x} \rangle}.
\end{equation}
Note that the choice of whether to use the left or right eigenstate to compute $E_\textrm{loc}$ does not matter since this formulation allows the Hamiltonian to produce the same energy when acting to the left or to the right, which we show in Sec. II of the Supplemental Material~\cite{supmat}. This formulation is particularly important in NH systems since it allows to bypass the direct implementation of the biorthogonal product, which will require to train two separate neural networks. Additionally, it is instructive to note that $|\Psi_{i,R}\rangle$ and $\langle \Psi_{j,L}|$
are not individually normalized and need not be, since the local energy in Eq.~\eqref{Eq4} is computed as a ratio of amplitudes, which remains invariant under global rescaling of the wavefunctions.

Before going into more detail on how to recover the true ground state of an NH system, it is important to first discuss what this means for an NH system. While some works provide a possible way to get to the ground state of an NH model \cite{yu2024non, chen2023non}, the concept of a ground state is, in principle, not well defined and may be subject to many interpretations due to the complex energy spectrum and the lack of a variational principle. In Hermitian systems, the ground-state wavefunction is the eigenstate corresponding to the lowest energy eigenvalue of the Hamiltonian, which is always real due to Hermiticity. In NH systems, the eigenvalues of the Hamiltonian can be complex valued, meaning there is no universally defined lowest energy in the conventional sense. Instead, the ground state can be defined in terms of eigenstates with the smallest real part, the smallest magnitude, the smallest imaginary part of the eigenvalues, or any other possible combination of these scenarios. Here, we propose to find the many-body wavefunction that minimizes the real part of the energy expectation value. We will see that the ground-state energy of the NH TFIM is, in fact, purely real without any imaginary contribution, as is also found in Refs.~\cite{lenke2021high, li2014conventional}.

An additional layer of complexity in the choice of the ground-state wavefunction is that the wavefunctions are handed (left, right), and one needs to choose which eigenstate to use in order to compute the local energy. The latter can be easily solved by using our formulation of the local energy in Eq.~\eqref{Eq4}, which becomes the true ground-state energy provided the corresponding ansatz (many-body eigenstate), $\Psi_{0, R/L}$, is the ground-state wavefunction of the Hamiltonian. This formulation then enables us not to care about whether we use the right or left wavefunction.

\begin{figure*}[ht!]
    \centering
    \subfloat[\centering ]{{\includegraphics[width=0.48\textwidth]{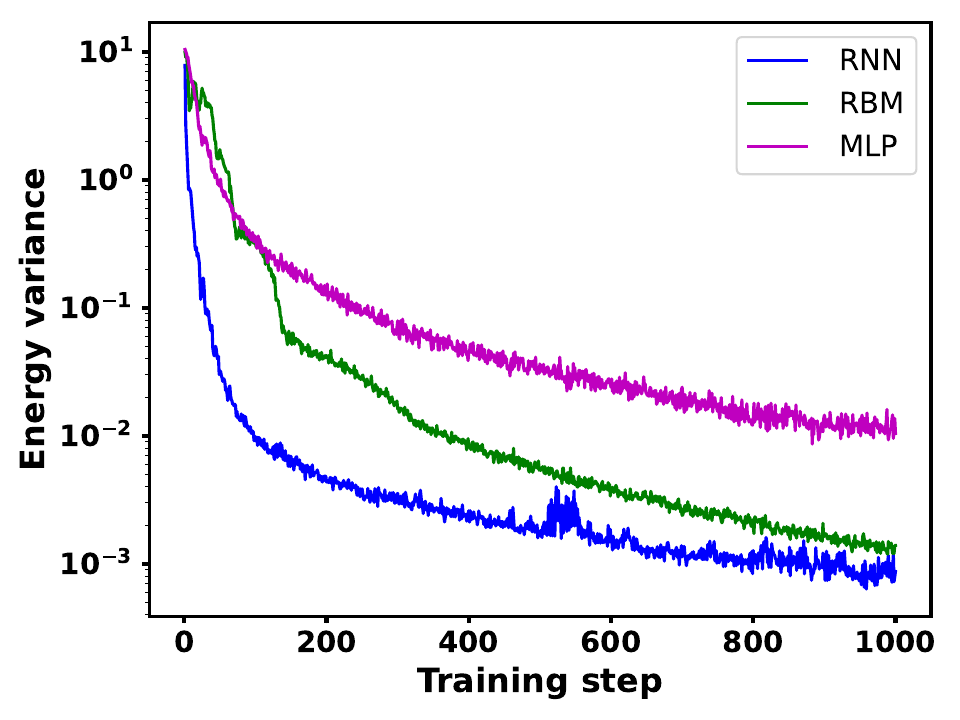} }}
    \subfloat[\centering ]{{\includegraphics[width=0.48\textwidth]{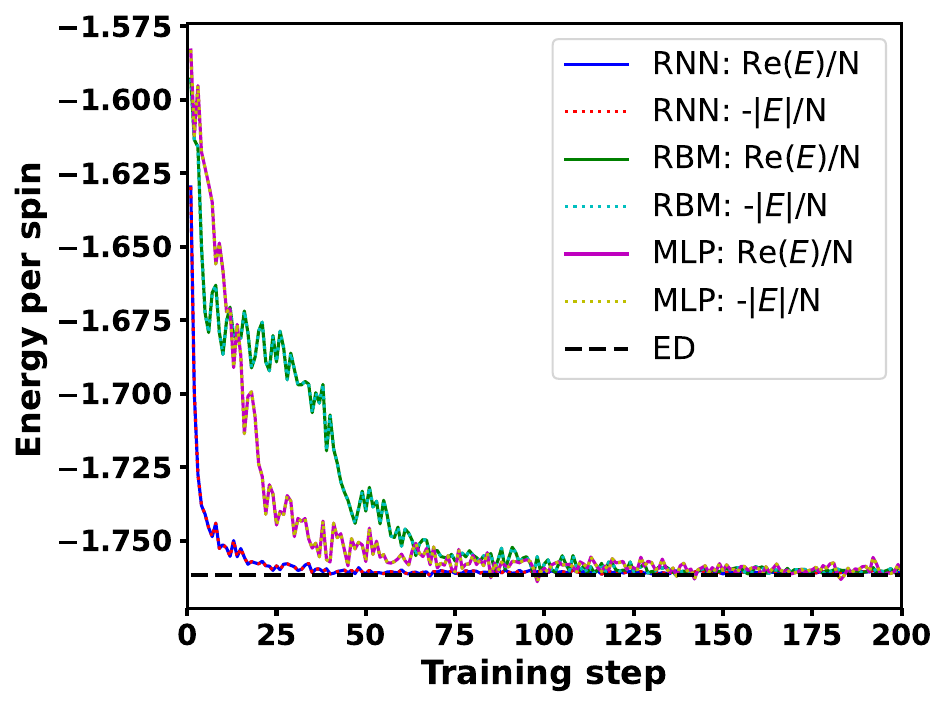} }}
    \caption{\textbf{Neural network architectures vs exact diagonalization}. First, we represent \textbf{(a)} the local energy variance of the real part of the average local energy for the RNN (blue), RBM (green), and MLP (magenta). The RNN converges faster than the RBM and the MLP. Then, we simulate \textbf{(b)} the ground-state energy of the model for the staggered magnetic field with the RNN (blue), RBM (green), and MLP (magenta). All neural network algorithms recover the true ground-state energy predicted by ED (black dashed line). We also plot the negative absolute value of the energy for the RNN (dashed red), RBM (dashed cyan) and, MLP (dashed yellow) for comparison. In both plots, we set $N=10$, $\eta =1.6$ and $\xi =0.16$. PBCs are considered in this figure.}
    \label{fig3}
\end{figure*}

\section{Ground state with neural networks}\label{S2}

Here, we use a neural network as an ansatz to a many-body wavefunction; that is, given an input configuration $\bm{x}$, it outputs $\Psi(\bm{x})$. We use the variational principle by using Monte Carlo sampling to sample several configurations $\bm{x}$ then computing and minimizing the average local energies given in Eq.~\eqref{Eq4} (see the Supplemental Material~\cite{supmat}). As we are looking for the state with lowest real energy, we use an algorithm that minimizes the real  part. We note, however, that our algorithm can be easily adapted to look for states with largest real part, largest or smallest imaginary part, largest or smallest absolute value, or even a certain combination of real and imaginary parts of the energy. As such, the algorithm can thus be readily adapted and applied to systems with a fully complex ground-state energy. Although our method is sufficiently general and robust to be applied to the strongly NH regime ($\eta = \xi$, and $\xi > \eta$ ) (see Supplemental Material~\cite{supmat}), we deliberately exclude the case $\xi > \eta$, as it corresponds to unrealistically large loss/gain rates that are both experimentally not of interest and of limited physical relevance. In this work, we therefore restrict our analysis to the weak NH limit ($\xi = \eta/10$), where NH effects remain appreciable without overwhelming the system's behavior. Throughout this paper, we set $\xi = \eta/10$ unless otherwise stated, and introduce a regularization factor $\alpha$ in the loss function to stabilize the optimization procedure and penalize large fluctuations introduced by the imaginary part. The role of $\alpha$ becomes especially crucial when focusing on eigenstates that have complex eigenvalues. In that case, we may want to focus on maximizing the imaginary part (long lifetime) or instead put constraints on the absolute value of the eigenenergies. This latter scenario is particularly relevant in transfer matrix methods and the generalized Brillouin zone problem~\cite{PhysRevB.104.104203,PhysRevB.99.245116,PhysRevLett.123.066404}, where the ordering of the absolute value of eigenvalues is of interest.

By minimizing the real part of energy, we identify the eigenstate that is energetically stable (lowest real part). As this state has a disappearing imaginary part, it is also dynamically stable. This eigenstate represents the ground state (in a generalized sense) of the NH system. Our approach acts as a physical filter that only focuses on the physically relevant spectrum instead of considering all states---many of which are noise in practice---as in the case of ED. A summarized general algorithm describing the aforementioned implementation for the RNN, RBM, and MLP can be found in the Supplemental Material~\cite{supmat}. It is worth emphasizing that this procedure is naturally also valid for the unbroken $PT$-symmetric regime, where all the local energies are real.

\subsection{Non-Hermitian many-body neural network wavefunctions}

Recent advances in machine learning have provided powerful tools for studying quantum many-body systems, particularly through neural network-based variational wavefunctions \cite{carleo2017solving, carrasquilla2021use, kim2024neural, hartmann2019neural, doschl2025neural, lange2024architectures, cai2018approximating}. Among these, RNNs \cite{hibat2020recurrent, hibat2024recurrent, medsker1999recurrent}, RBMs \cite{salakhutdinov2007restricted, melko2019restricted, hinton2012practical}, and MLPs \cite{murtagh1991multilayer, almeida2020multilayer, kruse2022multi, cai2018approximating} have emerged as effective representations for approximating quantum states. These architectures offer distinct advantages: RBMs are energy-based models made of a dense single-layer feed-forward network \cite{PhysRevB.107.195115} that capture nontrivial correlations with relatively few parameters, while MLPs provide a flexible framework for function approximation. RNNs, on the other hand, are particularly well-suited for sequential data and have demonstrated remarkable success in modeling quantum wavefunctions. A key motivation behind this framework is that this architecture enables autoregressive sampling. As a result, independent samples can be generated in a single network evaluation, eliminating the need for Markov chain Monte Carlo, which may suffer from long autocorrelation and thermalization times~\cite{reh2023optimizing}. By leveraging these methods, we aim to construct NH many-body wavefunctions and assess their ability to capture the system's critical behavior to simulate the ground-state properties of the model and compute the corresponding physical observables.

\begin{figure*}[ht!]
    \centering
    \subfloat[\centering ]{{\includegraphics[width=0.48\textwidth]{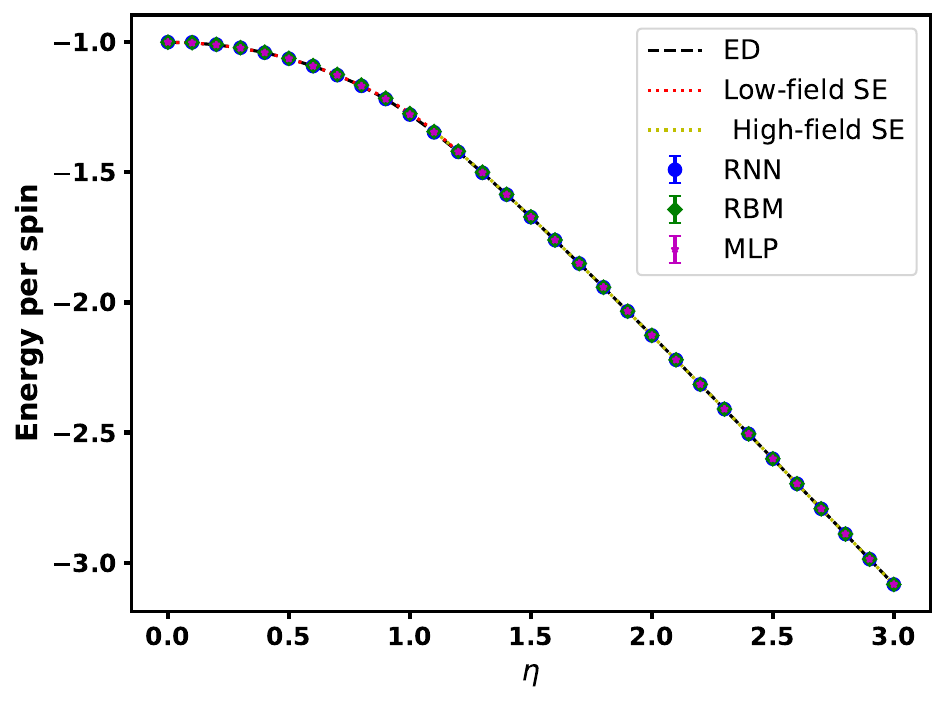} }}
    \subfloat[\centering ]{{\includegraphics[width=0.48\textwidth]{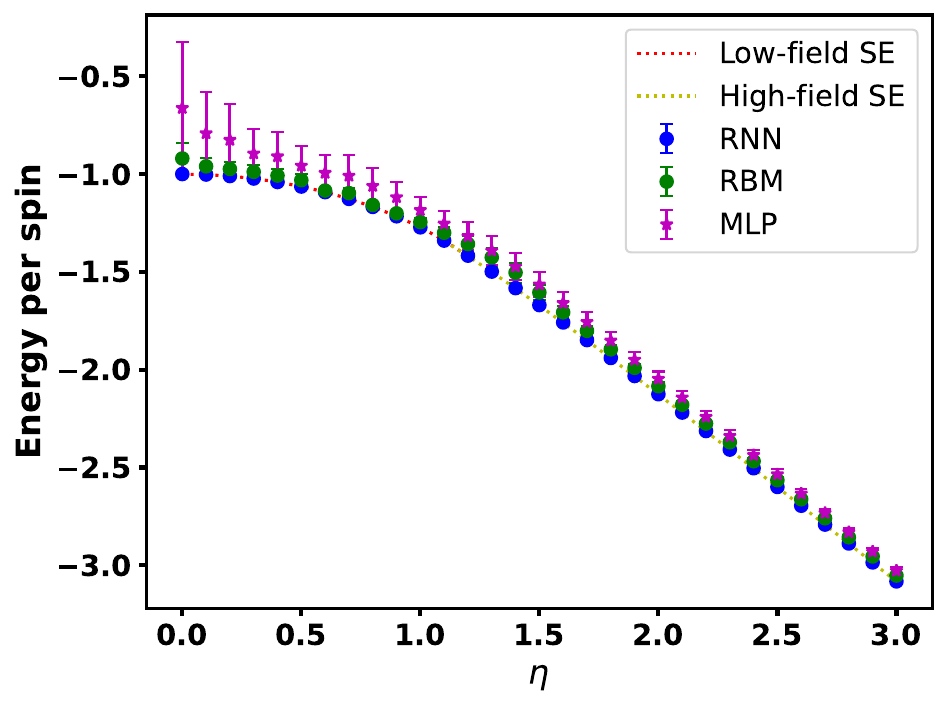} }}
    \caption{\textbf{Ground-state energy per spins}. We plot the average ground-state energy per spin against the $\eta$ for \textbf{(a)} the extrapolated high-order SE, where we depict both the low-field (dotted red) and the high-field (dotted yellow) regimes. We make use of an RNN (blue circles), an RBM (green circles), and an MLP (cyan stars), and compare the results with ED (dashed black line). The error bars are smaller than the data points and $N=10$. Beyond ED, for \textbf{(b)} $N=100$, we also compute the energy per spin for the RNN (blue circles), the RBM (green circles), and the MLP (magenta circles), and compare the result with the extrapolated higher-order SE (red and yellow dotted lines). The RNN is more efficient than the RBM and the MLP. $\xi=\eta/10$ for both figures. PBCs are considered in this figure.} 
    \label{fig6}
\end{figure*}

In this section, we simulate the ground state of the model using an RNN, an RBM, and an MLP. We benchmark our results against ED and the high-order series expansion (SE) derived in Ref.~\cite{lenke2021high} for the low- ($\eta/J, \xi/J \ll 1$) and high-field ($\eta/J, \xi/J \gg 1$) limits that we numerically extrapolated (see the Supplemental Material~\cite{supmat}) to capture the intermediate-field regime. In each of the neural network architectures, we enforced $PT$ symmetry on the many-body ansatz following the prescription in the Supplemental Material~\cite{supmat}. Although we enforce this symmetry on our original ansatz, the main idea is to have a neural network that can capture such a symmetry if present in the system at the beginning of the simulation [as it is the case when considering open boundary conditions (OBCs)]. However, it is instructive to note that the neural network breaks this symmetry as soon as we enter the broken regime fully consistent with the spontaneous breakdown of $PT$ symmetry for wavefunctions in this phase. A general overview of the algorithms and the parameters used for all simulations in this paper are summarized in the Supplemental Material~\cite{supmat}.

First, we plot and compare the energy variance for each architecture, as shown in Fig.~\ref{fig3}(a). The results indicate that the energy estimates exhibit low fluctuations and steadily converge toward a well-defined value for each architecture. We see that the RNN achieves a stable energy estimate more rapidly than both the RBM and MLP (as we in general expect for a Hermitian system as well), suggesting that it is more efficient in capturing the ground-state properties of the system. We also analyze the evolution of local energies during the training process for each architecture, [cf. Fig.~\ref{fig3}(b)]. As predicted by the variance, the RNN converges more rapidly. In the spontaneously broken symmetry regime, the local energies computed during training generally take complex values, but the final ground-state energy estimate is obtained through Monte Carlo averaging. Since these local energies predominantly appear in complex conjugate pairs, averaging leads to a cancellation of the imaginary components, yielding a real-valued result. This is effectively equivalent to taking the absolute value of the energy, as we confirm in Fig.~\ref{fig3}(b), where the negative sign indicates that the ground-state energy has a negative real part. The fact that the real part of the averaged energy closely matches its magnitude suggests that we are operating in a regime where only a few states spontaneously break $PT$ symmetry, placing the system near an exceptional point. Consequently, despite the broken symmetry at the local level, the global behavior exhibits ``real-like'' spectrum, consistent with proximity to the unbroken $PT$-symmetric phase. As a result, we are capturing the system's behavior close to an EP.

Furthermore, we investigate the ground-state energy per spin for each neural network architecture as a function of $\eta$ and construct a corresponding phase diagram (cf. Fig.~\ref{fig6}). In the small-system-size regime, where ED is feasible ($N=10$), all neural network architectures perform similarly, and their results closely match those obtained from ED and the extrapolated high-order SE as shown in Fig.~\ref{fig6}(a). However, for larger system sizes, beyond the reach of ED ($N = 100$), Fig.~\ref{fig6}(b), the autoregressive nature ~\cite{medsker2001recurrent, yu2019review} of the RNN allows it to effectively capture the system’s behavior, whereas the RBM and MLP struggle to achieve the same level of accuracy. This suggests that the RNN is robust and thus the most suitable architecture for our study (which is, in general, true for Hermitian models as well). Nevertheless, we later show that one can further improve the performances of all our neural network architectures using TL. 

\subsection{Recurrent neural network versus series expansion around criticality}

\begin{figure}[t] 
    \centering
\includegraphics[width=0.48\textwidth]{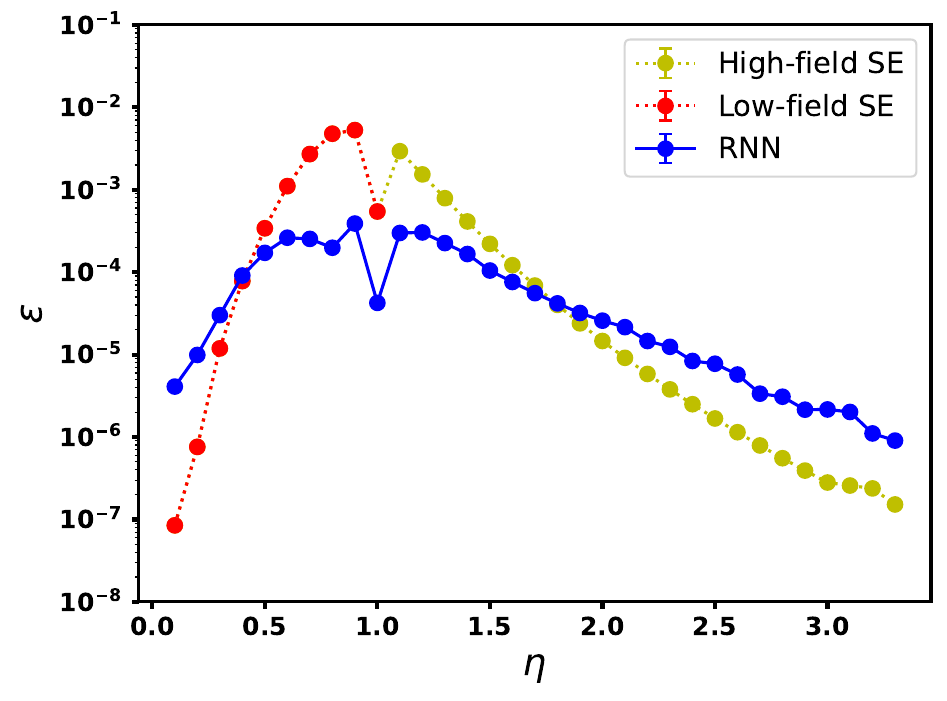}
    \caption{\textbf{Relative errors on RNN and SE}. We compute the relative errors $\varepsilon_\textrm{RNN}= |\textrm{Re}(E_\textrm{ED})-\textrm{Re}(E_\textrm{RNN})|/|\textrm{Re}(E_\textrm{ED})|$ (blue) and $\varepsilon_\textrm{SE}=|\textrm{Re}(E_\textrm{ED})-\textrm{Re}(E_\textrm{SE})|/|\textrm{Re}(E_\textrm{ED})|$ (red and yellow) for $N=10$ and $\xi=\eta/10$, where $E_\textrm{RNN}$ and $E_\textrm{SE}$ are the ground-state energies obtained with the RNN and the extrapolated SE, respectively. The error bars are smaller than the data points. PBCs are considered in this figure.}
    \label{fig7}
\end{figure}

In Ref.~\cite{lenke2021high}, Lenke \textit{et al.} employed the high-order SE to investigate the ground state in both the low- and high-field regimes with remarkable accuracy. Given that the RNN has demonstrated superior performance among the tested neural network architectures, we compare its accuracy near the transition point to that of the SE method. Our results indicate that, around this critical region, the RNN more effectively captures the system's behavior than the high-order SE approach. The relative error obtained with both methods is illustrated in Fig.~\ref{fig7}. This finding suggests that our algorithm is robust and that the RNN is a promising choice for studying NH quantum systems. 

In addition, an SE is inherently limited to perturbative corrections around a known limit (low- and high-field)~\cite{lenke2021high}. In contrast, an RNN-based, or any neural network-based, variational ansatz in general can be optimized globally, allowing it to capture the full phase diagram, including regimes where perturbation theory fails, e.g., near exceptional points or in strong anti-Hermitian fields. Also, the RNN allows efficient sampling of system configurations, which enables direct computation of physical observables, such as magnetization and entanglement entropy, across the entire phase diagram, whereas an SE typically focuses on specific quantities (like the ground-state energy and the energy gap) and only in particular limits. Furthermore, the high-field gap in SE is inaccessible perturbatively due to complex energy eigenvalues and the presence of exceptional lines~\cite{lenke2021high}. Since an RNN-based approach works by direct optimization of a trial wavefunction, it can learn the structure of eigenstates even in these non-trivial regions. Moreover, the RNN ansatz (gated recurrent units~\cite{cho-etal-2014-properties} or the long short-term memory units~\cite{hochreiter1997long}) is particularly powerful for capturing long-range correlations in 1D quantum systems~\cite{hibat2020recurrent, supmat}. This is crucial for $PT$-symmetric models, where NH effects can induce complex correlation patterns that standard perturbative methods may not fully describe. So, the RNN implementation provides a more complete picture of $PT$-symmetric phase transitions beyond what the SE method can achieve and offers much more flexibility. Nevertheless, the SE outperforms the RNN at high- and low-field limits (see Fig.~\ref{fig7}), and allows to access the functional approximate form of the ground-state energy.

\begin{figure*}[!ht]
    \centering
    \subfloat[\centering ]{{\includegraphics[width=0.33\textwidth]{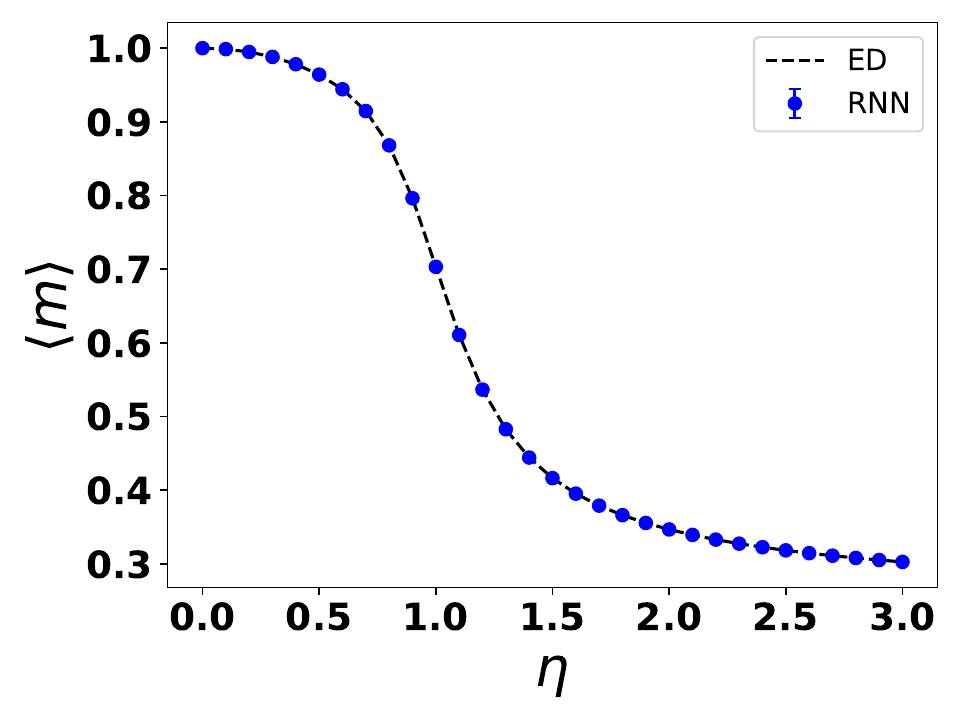} }}
    \subfloat[\centering  ]{{\includegraphics[width=0.33\textwidth]{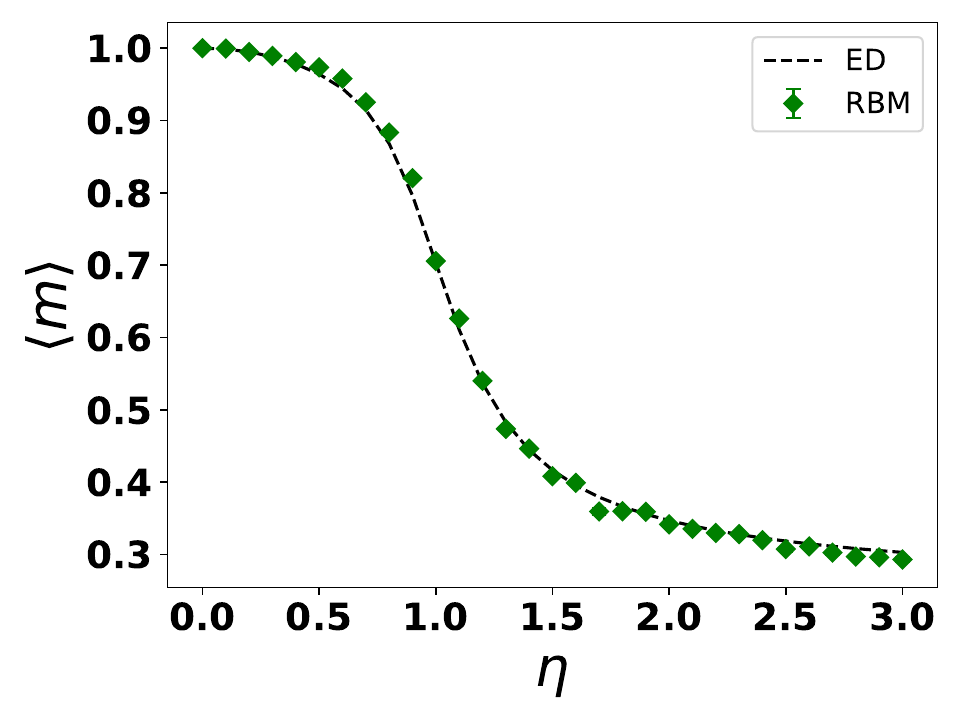} }}
    \subfloat[\centering  ]{{\includegraphics[width=0.33\textwidth]{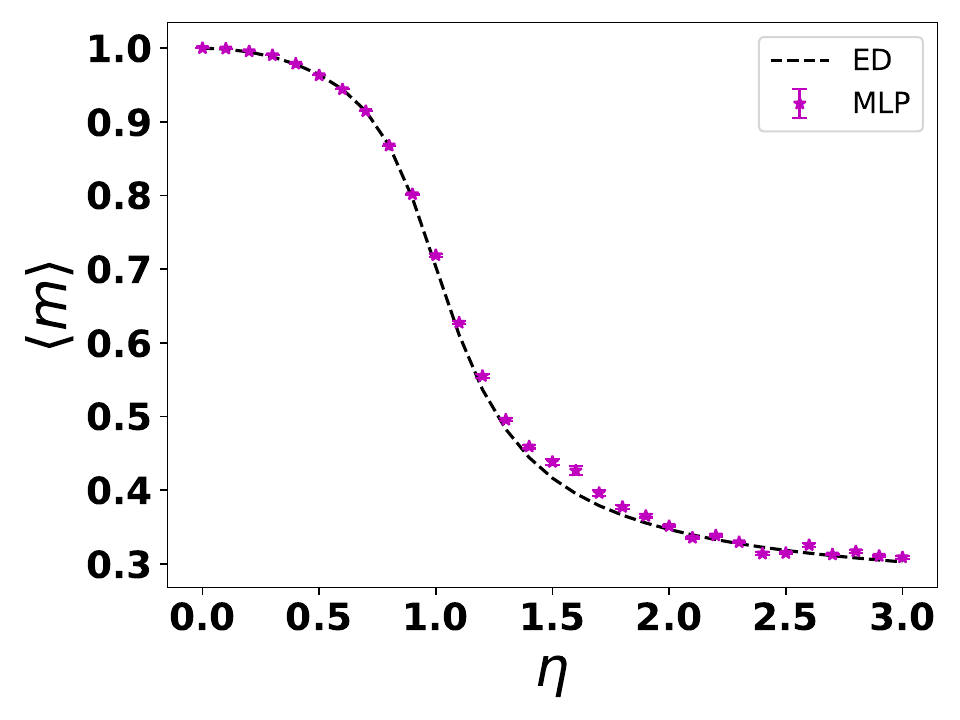} }}
    \qquad
    \subfloat[\centering ]{{\includegraphics[width=0.33\textwidth]{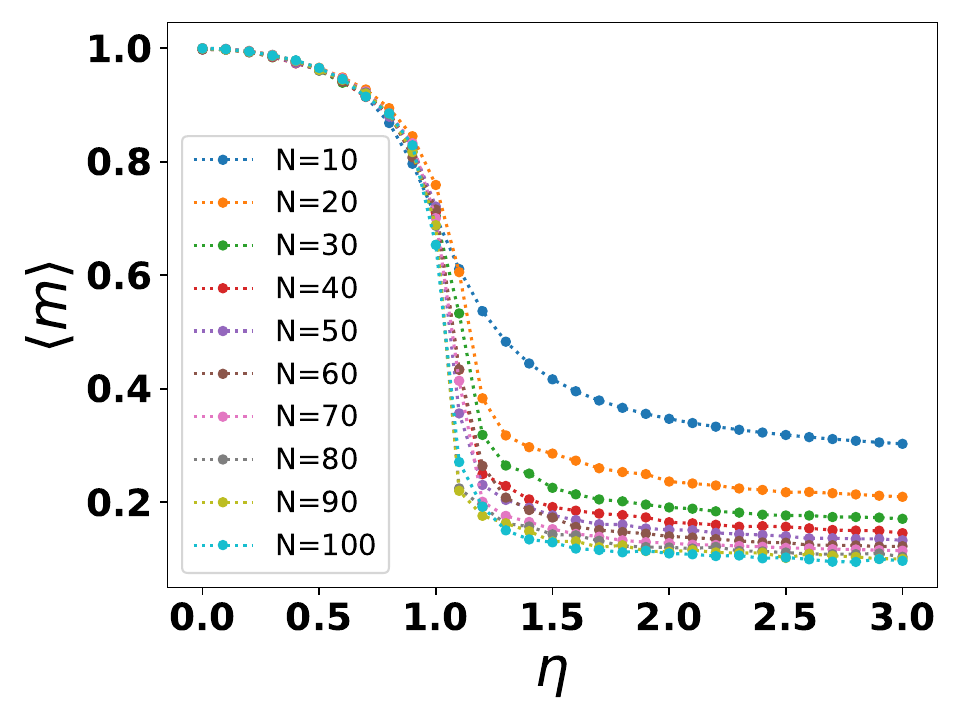} }}
    \subfloat[\centering ]{{\includegraphics[width=0.33\textwidth]{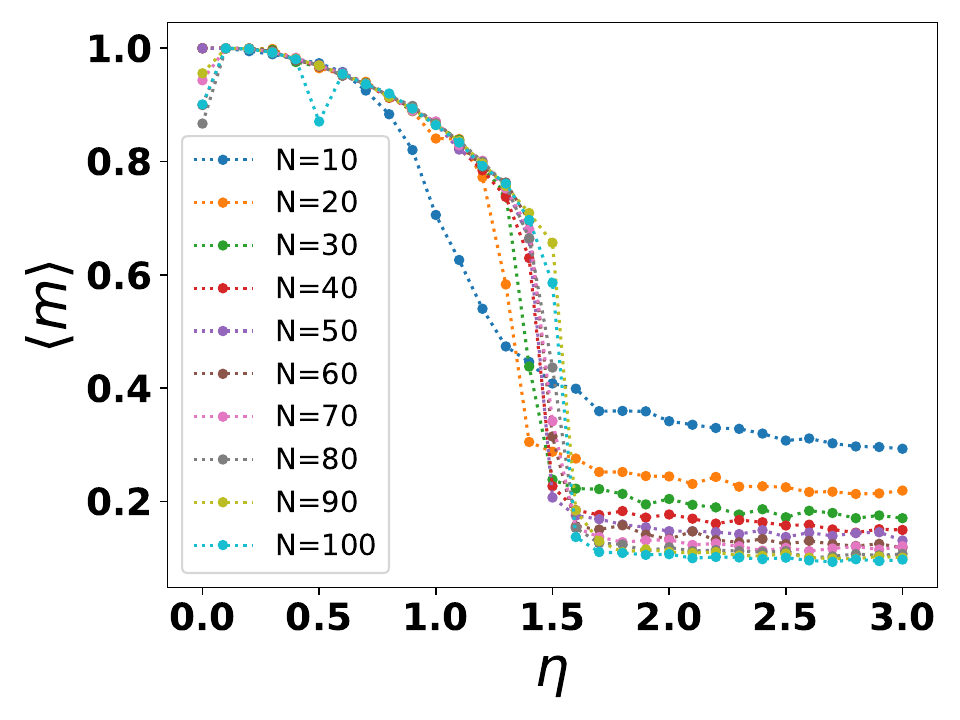} }}
    \subfloat[\centering  ]{{\includegraphics[width=0.33\textwidth]{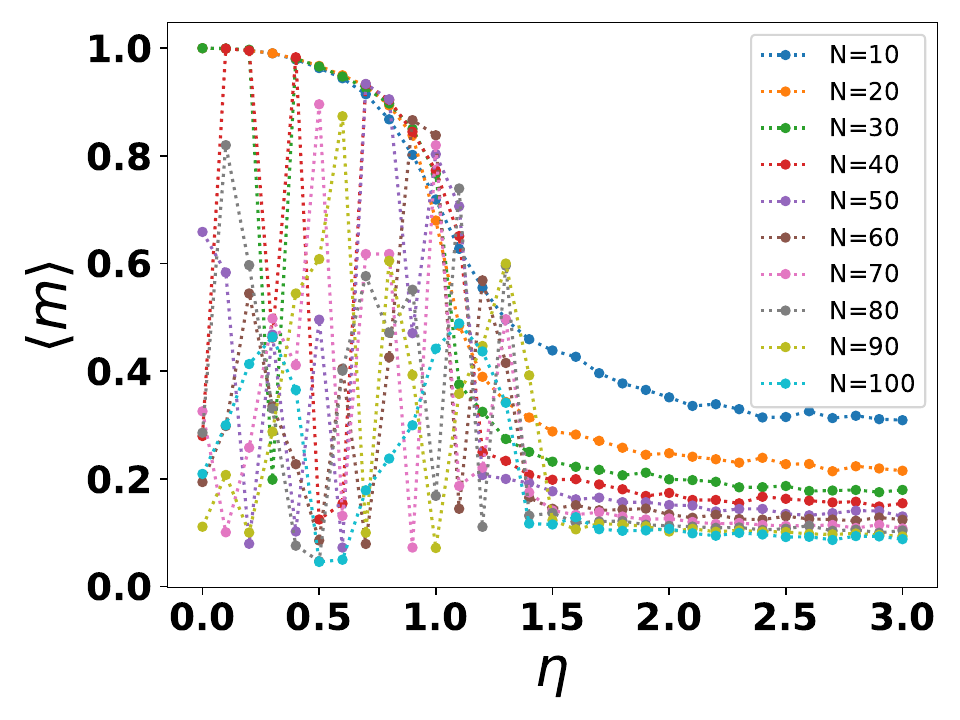} }}
    
    \qquad
    \subfloat[\centering  ]{{\includegraphics[width=0.33\textwidth]{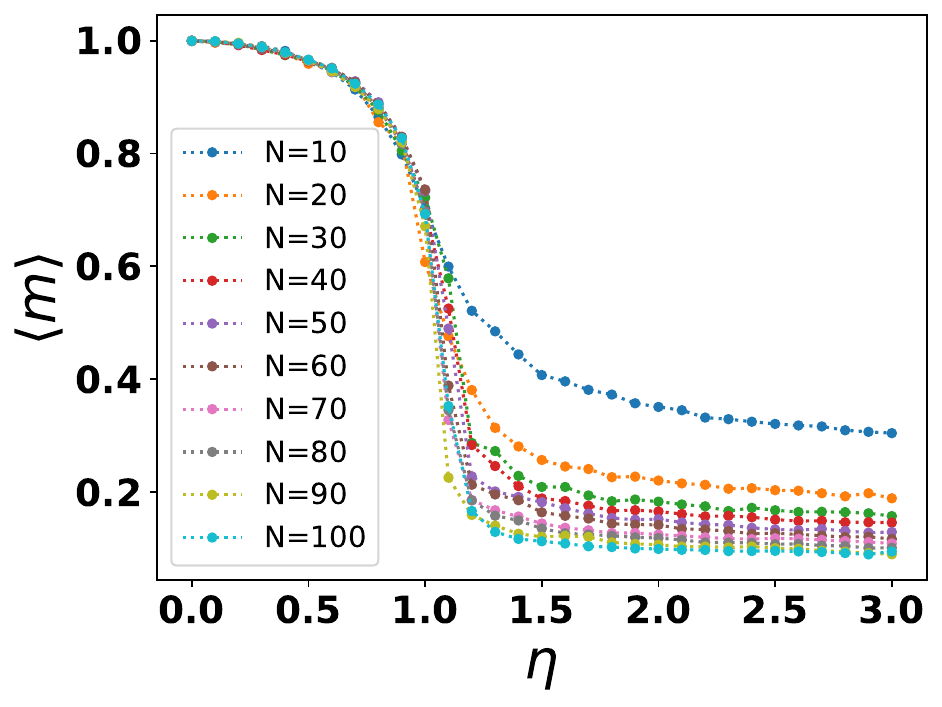} }}
    \subfloat[\centering   ]{{\includegraphics[width=0.33\textwidth]{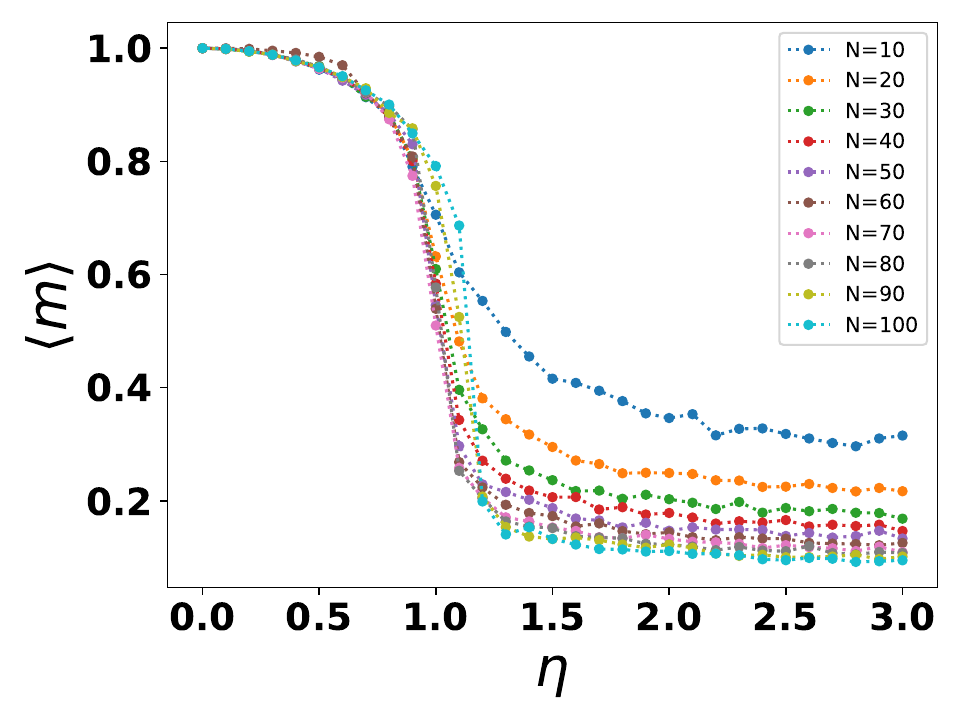} }}
    \subfloat[\centering  ]{{\includegraphics[width=0.33\textwidth]{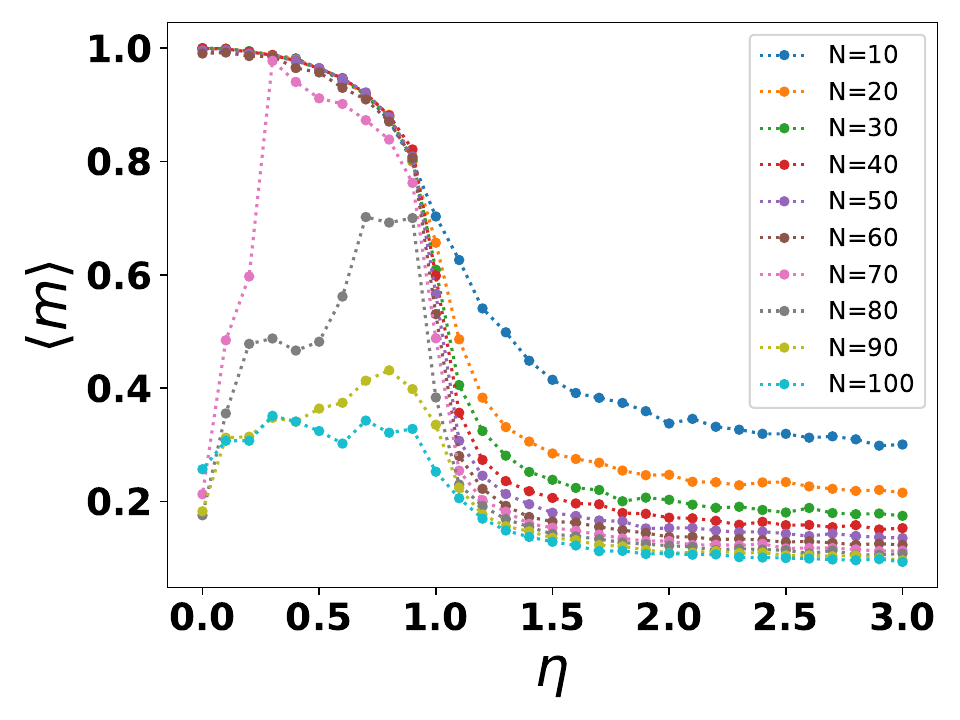} }}
    
    \qquad
    \subfloat[\centering ]{{\includegraphics[width=0.33\textwidth]{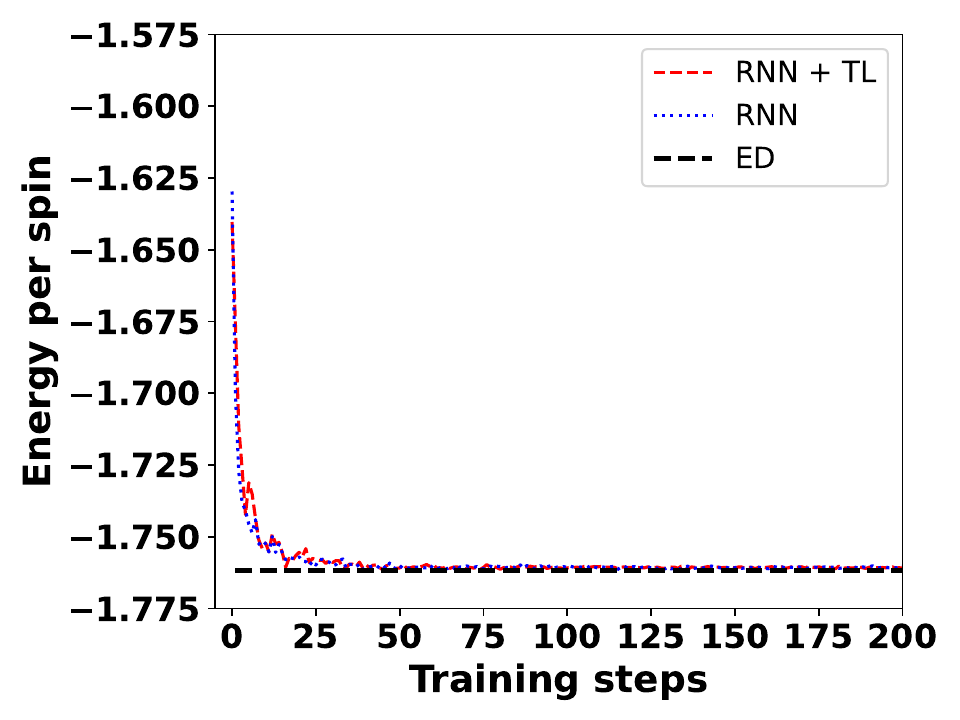} }}
    \subfloat[\centering   ]{{\includegraphics[width=0.33\textwidth]{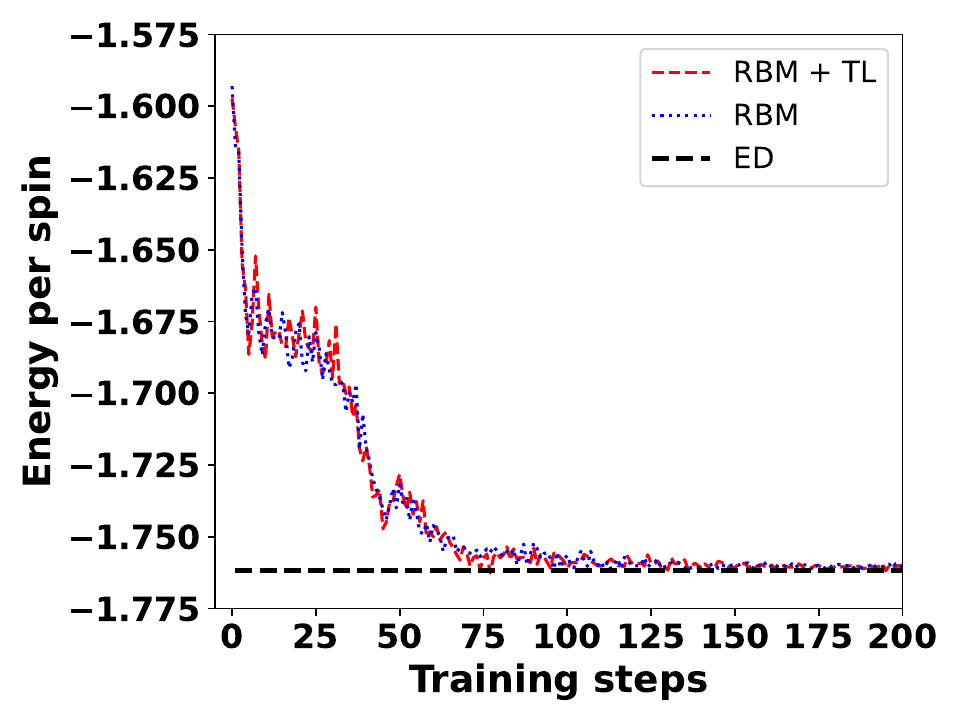} }}
    \subfloat[\centering   ]{{\includegraphics[width=0.33\textwidth]{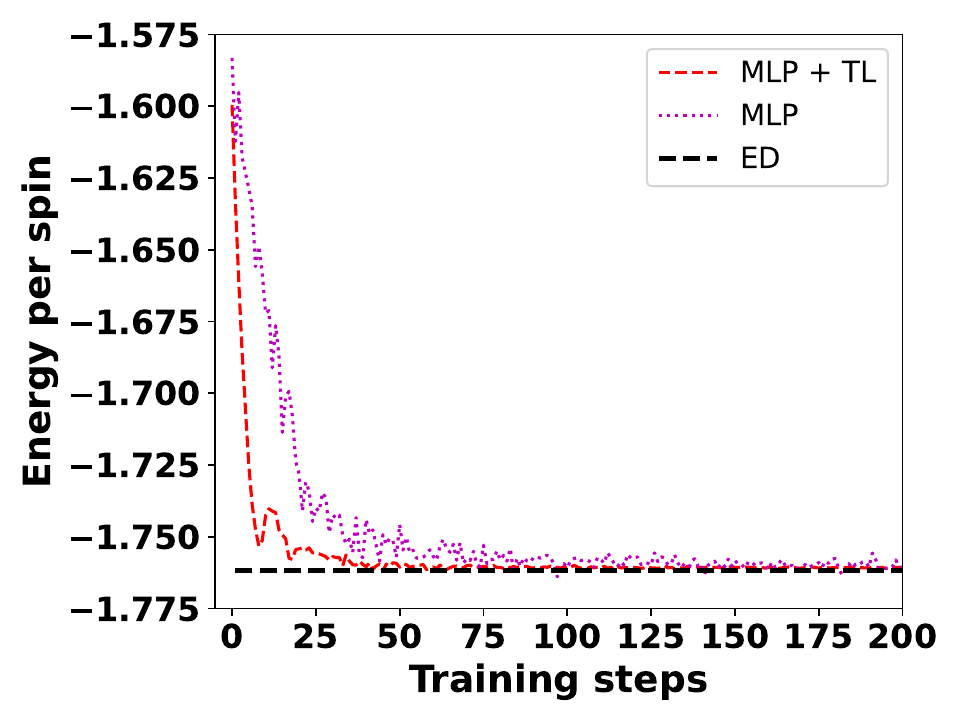} }}

    \caption{\textbf{Transfer learning}. We plot the magnetization of the system $\left<m\right>$ for different system sizes across various neural network architectures. The first column corresponds to simulations using the RNN, while the second and third columns represent results from the RBM and MLP, respectively. The first row [panels \textbf{(a)-(c)}] shows the magnetization for $N = 10$ and $\xi =0.16$ compared to ED. The second row [panels \textbf{(d)-(f)}] presents the magnetization for different system sizes for each architecture, while the third row [panels \textbf{(g)-(i)}] incorporates transfer learning to enhance performance. The final row [panels \textbf{(j)-(l)}] illustrates the training process for each architecture, both with and without transfer learning for $N = 10$, $\eta =1.6$, and $\xi =0.16$ compared to ED. The quantum phase transition occurs at the critical value $\eta_c \approx 0.9871(1)$. The critical value $\eta_c$ shifts slightly as a response to non-Hermiticity rather than from numerical inaccuracies. PBCs are considered in this figure. }
    \label{fig8}
\end{figure*}

\section{Transfer Learning and Quantum Phase Transition }\label{S3}
Previously, we analyzed the behavior of the energy per spin as a function of $\eta$ for different system sizes. Building on this, we now turn our attention to phase transitions in magnetization, $\left<m\right> =N^{-1}\sum_{i=1}^N \langle \sigma_i^z \rangle$, which serves as an order parameter. Additionally, we aim to enhance the efficiency of our NQS using TL.

Humans naturally transfer knowledge between tasks by recognizing and applying relevant information from past experiences to unfamiliar situations. The closer a task is to something we have learned before, the easier it is to master. This principle is the same as the one used in TL~\cite{torrey2010transfer}, in contrast to traditional machine learning algorithms, which typically focus on isolated tasks. TL aims to bridge this gap by developing methods that leverage knowledge from one or more source tasks to enhance learning in a related target task. From a physical perspective, one would anticipate a connection between the wavefunctions of systems that share the same parameter values but differ in size, as if they represent the same system observed at varying length scales \cite{zen2020transfer,zen2020finding,rende2024fine}. Here, we investigate TL strategies inspired by this principle, and show that they outperform a ``cold-start'' approach in terms of both effectiveness and efficiency. 

First, we compute the magnetization within ED, where all neural network architectures perform relatively well, though the RBM and MLP exhibit some errors as shown in Figs.~\ref{fig8}(a)-ref{fig8}(c). For larger system sizes ($N = 100$), only the RNN recovers the magnetization. However, for each system size, the algorithm must be ``cold-started'' making the simulations computationally expensive. To address this, we propose reusing the same samples and weights from the initial value of  $\eta$ (i.e, $\eta =3.0$) to simulate the system at a later value of $\eta$ (that is $\eta =2.9$) (see Fig.~\ref{fig8}), and then fine tune the neural network model. This process is repeated until $\eta =0.0$.

By employing TL, we first show that the performance of both the RBM and MLP improves substantially, enabling the simulation of larger system sizes [see Figs.~\ref{fig8}(h), and ~\ref{fig8}(i)]. Nevertheless, even with TL, the MLP exhibits poor performance for $\eta < 1$, and system sizes $N =\in [70, 100]$ [see Figs.~\ref{fig8}(i)]. The pronounced fluctuations in the MLP results  can be traced to the architectural limitations of feedforward networks in representing quantum states with strong local correlations. In contrast to RNNs and RBMs — which are intrinsically suited to capture sequential and local dependencies in one-dimensional systems — MLPs treat the input as an unstructured vector and lack inductive bias with respect to the lattice topology. As a consequence, they are more susceptible to optimization noise and exhibit less stable convergence, particularly in NH settings where the energy landscape is highly intricate. Furthermore, TL enhances the efficiency of all architectures, making the simulations faster and more effective, cf. Figs.~\ref{fig8}(j)-\ref{fig8}(l). It can be seen that the energy per spin with transfer learning always starts lower than the energy per spin without transfer learning. This improvement is particularly evident for the MLP and RBM, where the TL-based algorithm converges more rapidly. In general, incorporating TL across all architectures reduces the initial computational cost, as the algorithm benefits from the prior knowledge encoded in the reused weights and biases. The simulation time for these simulations can be found in the Supplemental Material~\cite{supmat}.

In the standard Hermitian TFIM, the phase transition occurs between the ordered (ferromagnetic) phase, where the spins align and lead to a non-zero magnetization, and the disordered (paramagnetic) phase, where the spins become disordered and the magnetization approaches zero~\cite{pfeuty1970one}. In our case, the situation is more complex due to the $PT$ symmetry, and the phase transition could also refer to an abrupt change in the eigenstates of the model \cite{li2014conventional}; i.e., they may coalesce, as we will discuss more in detail later. However, the magnetization can still serve as a signature of the ordering or quantum phase transition that is the transition between the ordered and disordered phases. This allows us to locate the critical point where this transition occurs ($\eta_c \approx 0.9871(1)$) as shown in Figs.~\ref{fig8}(a)-\ref{fig8}(i). This critical value $\eta_c$ shifts slightly as a response to non-Hermiticity.

\section{Conclusion}\label{S5}

In this work, we demonstrate how conventional machine learning and artificial intelligence techniques can be extended to study NH systems. We go beyond previous studies using neural networks to study NH systems by considering a model with a complex spectrum. To study these models, we explicitly address the challenge posed by biorthogonality through formulating the local energy in a way that is independent of whether one works with the left or right eigenstate. We then show that for small system sizes, the RNN, RBM, and MLP architectures successfully recover the ground state with comparable accuracy. Their performance is benchmarked against ED and a high-order SE extrapolation, which we use to capture the system's behavior in the intermediate field regime. Despite the high accuracy of SE, we find that near criticality, the RNN demonstrates superior performance as one will expect (in general) for a Hermitian system. For larger system sizes, the RNN outperforms both the RBM and MLP, while achieving accuracy comparable to SE. However, unlike SE, the RNN allows for efficient computation of physical observables. Moreover, we study the magnetic phase transitions and show that employing TL enhances the performance of all neural network architectures. In particular, TL significantly accelerates the convergence of RBM and MLP, enabling them to compute magnetization at larger system sizes with improved accuracy. Finally, we use ED to investigate the unbroken-to-broken phase transition of the system under OBCs and the emergence of EPs. We extend the technique based on computing the maximum overlap of the ``all-to-all'' dot product between eigenstates used in Refs.~\cite{Mandal2021, Montag2024} to many-body NH systems and enable the identification of EPs.

We note that the algorithm developed in this work can be straightforwardly adapted to find different states. While we are interested in finding the state with the lowest real energy in this work, one may instead be interested in finding, e.g., excited states such as the state with the largest imaginary energy. In this case, instead of defining the loss function $\mathcal{L}$ in terms of the real part of the eigenvalues, one can define it in terms of the imaginary parts. Indeed, the $n$-th excited state $\psi_n$ can be constructed from spin configurations that are biorthogonal to all lower-energy states, $\{\psi_0, \psi_1, \dots, \psi_{n-1}\}$, ensuring sequential biorthogonality across the spectrum.

Another extension of this work is to study systems beyond non-interacting $PT$-symmetric models. While our method is tailored here to focus on the real part of the spectrum, it can be straightforwardly tailored to take the complete complex picture into account. The main challenge for such an extension actually lies in the ansatz of the many-body wavefunction, which should then, in principle, obey a different symmetry. Additionally, adding interactions or moving to higher dimensions would provide another avenue to test our machinery. In higher dimensions, the single hidden state propagated between sites can be replaced by multiple hidden states~\cite{PhysRevB.109.245120}, each corresponding to a neighboring site in a different spatial direction, with the recursion relation for the hidden states modified accordingly to reflect the lattice geometry. The training procedure can follow the approach of Ref.~\cite{solinas2025biorthogonal}, where the loss function is formulated as a biorthogonal product of the left and right eigenstates. However, this prescription can be simplified in the presence of symmetries. For instance, in $PT$-symmetric systems, the left and right eigenstates are related by complex conjugation, allowing the algorithm to proceed similarly to that of a Hermitian system.

\section{Acknowledgments}

We thank Kai Phillip Schmidt for insightful discussion on the model Hamiltonian. We thank Anton Montag for the discussions on confirming the emergence of exceptional points in our system. We also thank Jan Alexander Koziol for enriching discussions on the numerical implementation of the local energy. L.W. and F.K.K. acknowledge funding from the Max Planck Society Lise Meitner Excellence Program~\mbox{2.0}. F.K.K. also acknowledges funding from the European Union's ERC Starting Grant ``NTopQuant'' (101116680). Views and opinions expressed are, however, those of the authors only and do not necessarily reflect those of the European Union or the European Research Council (ERC). Neither the European Union nor the granting authority can be held responsible for them.

\section{Data Availability}

The data that support the findings of this article are openly
available~\cite{GitHubRepo}; embargo periods may apply.

\appendix

\section{Effect of Boundary Conditions}\label{A5}

\begin{figure*} 
    \centering
    \subfloat[\centering ]{{\includegraphics[width=0.48\textwidth]{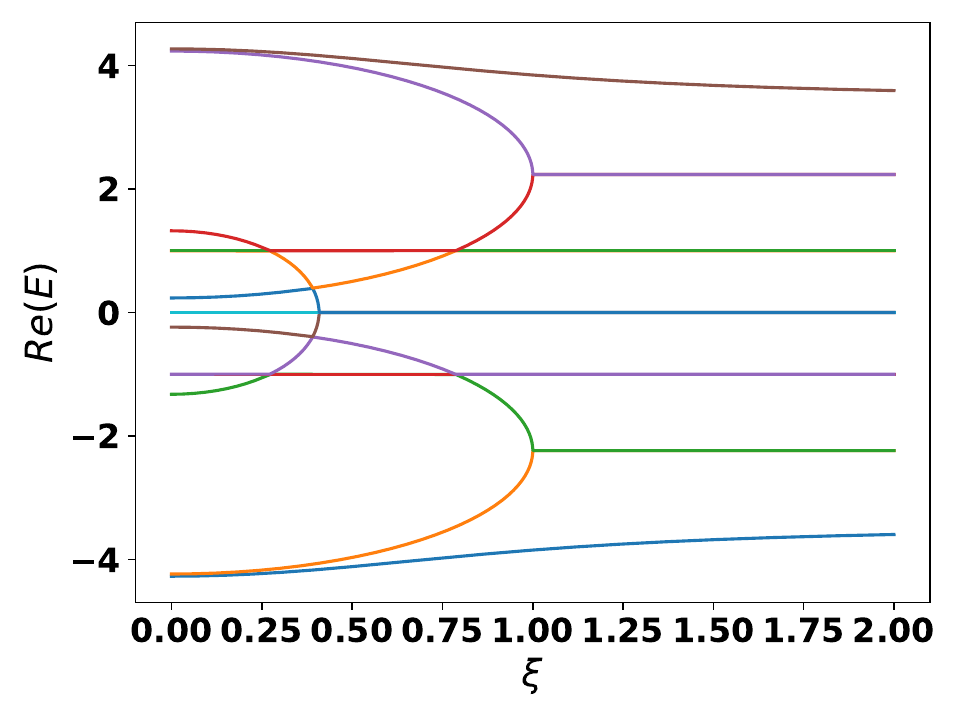} }}
    \subfloat[\centering ]{{\includegraphics[width=0.48\textwidth]{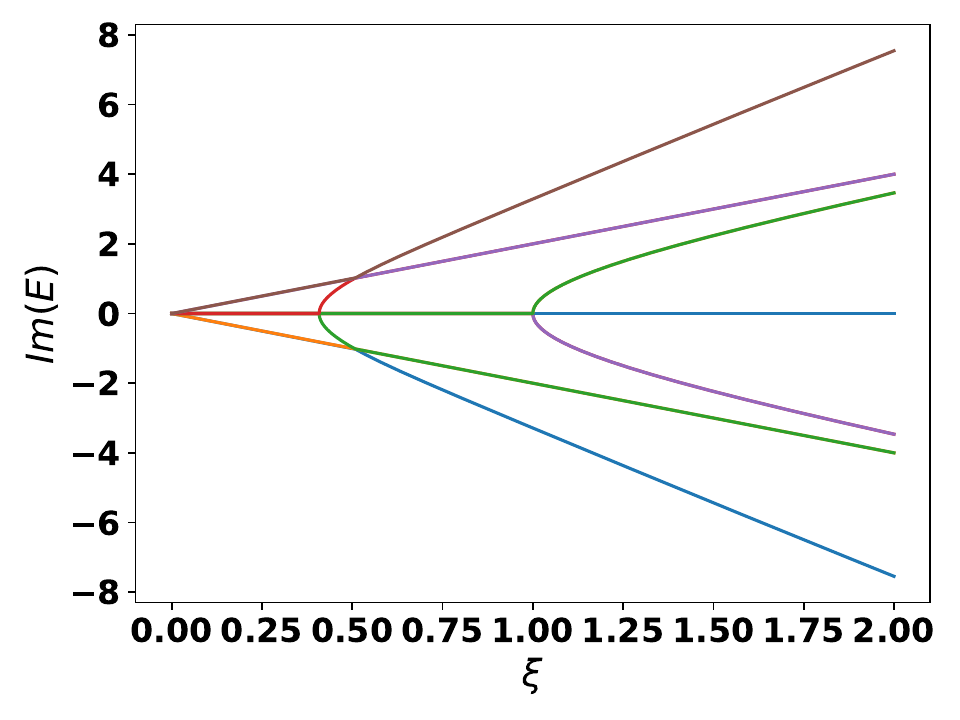} }}
    \caption{\textbf{Energy spectrum under PBCs}. We plot the \textbf{(a)} real and the \textbf{(b)} imaginary parts of the ground state energy under PBCs for $\eta = 0.5$. One enters the broken $PT$-symmetric regime as soon as one turns on the non-Hermiticity, $\xi \neq 0$. We consider $N=4$ and the coloring is  determined by the order in which the bands appear.}
    \label{fig20}
\end{figure*}

The analysis presented in Secs.~II to IV in the main text was conducted considering PBCs. As predicted in Ref.~\cite{lenke2021high}, we show here that $PT$ symmetry is spontaneously broken by the eigenstates as soon as one turns on the non-Hermiticity ($\xi \neq 0$) under PBCs (see Fig.~\ref{fig20}) while the Hamiltonian remain $PT$-symmetric. In contrast, Fig.~7 in the main text shows that under open boundary conditions (OBCs), this phase transition is not immediate and only occurs later at some particular values of the NH perturbation.

\section{Local Energy and Algorithms}\label{A3}

\subsection{Local Energy}

The local energy, $E_\textrm{loc}(\bm{x})$ in Eq.~(4) in the main text with $\bm{x}$ being the spin configuration, plays a crucial role in the VMC methods, as it provides an efficient way to estimate the expectation value of the Hamiltonian $H$ in a given quantum state. It is defined as the ratio of the Hamiltonian’s action on the wavefunction to the wavefunction itself, evaluated at a specific configuration $\bm{x}$. 

In non-Hermitian systems, where the left and right eigenstates of the Hamiltonian are distinct, the local energy can be computed using either the right eigenstate  $|\Psi_{0,R}\rangle$ or the left eigenstate $\langle\Psi_{0,L}|$, as shown in  Eq.~(4) in the main text. Importantly, despite the NH nature of the system, this formulation ensures that the energy remains consistent, whether the Hamiltonian acts from the left or the right. This property makes our formulation of the local energy particularly useful for numerical simulations, as it allows for a meaningful energy evaluation even when conventional Hermitian approaches fail. From the biorthogonality relation in Eq.~(2) in the main text, the ground-state energy of the Hamiltonian is
\begin{equation}\label{EqA1}
    E_0 =\langle H \rangle_0 = \frac{\langle \Psi_{0,L}|H|\Psi_{0,R}\rangle}{\langle \Psi_{0,L}| \Psi_{0,R}\rangle}.
\end{equation}
By using the completeness of the spin basis and the fact that $\Psi_R(\bm{x}) = \langle \bm{x}| \Psi_R \rangle$ or $\Psi_L(\bm{x}) = \langle \Psi_L | \bm{x} \rangle$  one finds
\begin{align} \label{EqA2}
    \langle H \rangle &= \frac{\sum_{\bm{x}}\langle \Psi_{0,L}|\bm{x}\rangle \langle \bm{x}|H|\Psi_{0,R}\rangle}{\sum_{\bm{x}}\langle \Psi_{0,L}|\bm{x}\rangle \langle \bm{x}| \Psi_{0,R}\rangle} \notag \\
    &= \frac{ \sum_{\bm{x}} \Psi_{0,L}(\bm{x}) \left[ \frac{\langle \bm{x}|H|\Psi_{0,R}\rangle}{\langle \bm{x}| \Psi_{0,R}\rangle} \right] \Psi_{0,R}(\bm{x})}{\sum_{\bm{x}} \Psi_{0,L}(\bm{x})\Psi_{0,R}(\bm{x})}.
\end{align}
The local energy is then taken to be
\begin{equation}\label{EqA3}
    E_\textrm{loc}(\bm{x})=\frac{\langle\bm{x}|H|\Psi_{0,R}\rangle}{\langle \bm{x}| \Psi_{0,R}\rangle},
\end{equation}
where the summation of these energies becomes the true ground-state energy provided a good approximation of the ansatz. The expression of $E_\textrm{loc}(\bm{x})$ in terms of $\langle \Psi_{0,L}|$  can be found in a similar fashion.

\subsection{Algorithms}

\vspace{2mm} 
\hrule
\vspace{2mm}
\textbf{Recurrent Neural Network (RNN) training}
\vspace{1mm}
\hrule
\vspace{1mm}
\begin{algorithmic}[1]
\State Initial random spin configuration: $\bm{x}$
\State Initial random RNN ansatz: $rnn$
\For{$\bm{it}$ in range($\bm{steps}$) }
    \State $\bm{x} = $ \textit{Autoregressive sampling with RNN} 
    \State $\log \Psi(\bm{x}) = rnn(\bm{x})$
    \State $E_\textrm{loc}(\bm{x}) = compute\_eloc(\bm{x}, \log \Psi(\bm{x}))$ with Eq.~\eqref{EqA3}
    \State $loss\_real = \langle\log \Psi(\bm{x})Re(E_\textrm{loc}(\bm{x})) - \log \Psi(\bm{x})\langle Re(E_\textrm{loc}(\bm{x})) \rangle\rangle$
    \State $grad =gradient(loss\_real, rnn)$
    \State Update $rnn$ parameters with gradient descent.
\EndFor
\end{algorithmic}
\vspace{1mm}
\hrule
\vspace{1mm}

\vspace{2mm} 
\hrule
\vspace{2mm} 
\textbf{Restricted Boltzmann Machine (RBM) training}
\vspace{1mm}
\hrule
\vspace{1mm}
\begin{algorithmic}[1] 
\State Initial random spin configuration: $\bm{x}$
\State Initial random RBM ansatz: $rbm$
\For{$\bm{it}$ in range($\bm{steps}$) }
    \State $\bm{x} = $ \textit{Gibbs sampling with RBM} 
    \State $\log \Psi(\bm{x}) = rbm(\bm{x})$
    \State $E_\textrm{loc}(\bm{x}) = compute\_eloc(\bm{x}, \log \Psi(\bm{x}))$ with Eq.~\eqref{EqA3}
    \State $loss\_real = \langle\log \Psi(\bm{x})Re(E_\textrm{loc}(\bm{x})) - \log \Psi(\bm{x})\langle Re(E_\textrm{loc}(\bm{x})) \rangle\rangle$
    \State $grad =gradient(loss\_real, rbm)$
    \State Update $rbm$ parameters with gradient descent.
\EndFor
\end{algorithmic}
\vspace{1mm} 
\hrule
\vspace{1mm}

\vspace{2mm} 
\hrule
\vspace{2mm} 
\textbf{Multilayer Perceptron (MLP) training}
\vspace{1mm} 
\hrule
\vspace{1mm}
\begin{algorithmic}[1] 
\State Initial random spin configuration: $\bm{x}$
\State Initial random MLP ansatz: $mlp$
\For{$\bm{it}$ in range($\bm{steps}$) }
    \State $\bm{x} = $ \textit{Metropolis sampling with MLP} 
    \State $\log \Psi(\bm{x}) = mlp(\bm{x})$
    \State $E_\textrm{loc}(\bm{x}) = compute\_eloc(\bm{x}, \log \Psi(\bm{x}))$ with Eq.~\eqref{EqA3}
    \State $loss\_real = \langle\log \Psi(\bm{x})Re(E_\textrm{loc}(\bm{x})) - \log \Psi(\bm{x})\langle Re(E_\textrm{loc}(\bm{x})) \rangle\rangle$
    \State $grad =gradient(loss\_real, mlp)$
    \State Update $mlp$ parameters with gradient descent.
\EndFor
\end{algorithmic}
\vspace{1mm}  
\hrule
\vspace{1mm}

\section{Extrapolation of the Series Expansion}\label{A4}

The ground-state energy per spin of the NH TFIM at low- ($\eta/J, \xi/J \ll 1$) and high-field ($\eta/J, \xi/J \gg 1$) was derived in Ref.~\cite{lenke2021high}, and is given by

\begin{align} 
E_\textrm{low}^{(\eta, \xi)} &= -J - \frac{\eta^2 - \xi^2}{4J} - \frac{\eta^4 + \xi^4}{64J^3} - \frac{5 \eta^2 \xi^2}{32J^3} - \frac{\eta^6 - \xi^6}{256J^5} \notag \\
&- \frac{7 (\eta^4 \xi^2 - \eta^2 \xi^4)}{256 J^5}  
- \frac{25 (\eta^8 + \xi^8)}{16384J^7} -\frac{129 (\eta^6 \xi^2 + \eta^2 \xi^6)}{4096J^7}\notag \\
&- \frac{171 \eta^4 \xi^4}{8192J^7} 
- \frac{49 (\eta^{10} - \xi^{10})}{65536J^9} + \frac{781 (\eta^8 \xi^2 - \eta^2 \xi^8)}{65536J^9} \notag \\ 
&- \frac{33 (\eta^6 \xi^4 - \eta^4 \xi^6)}{32768J^9} - \frac{441 (\eta^{12} + \xi^{12})}{1048576J^{11}} \notag \\
&-\frac{7631 (\eta^{10} \xi^2 + \eta^2 \xi^{10})}{524288J^{11}}- \frac{18551 (\eta^8 \xi^4 + \eta^4 \xi^8)}{1048576J^{11}}  \notag \\
&- \frac{7241 \eta^6 \xi^6}{262144J^{11}}, \end{align}\label{d1}
and
\begin{align}
E_\textrm{high}^{(\eta, \xi)} &= -\eta - \frac{J^2}{4\eta} - \frac{J^4 (\eta^2 - 3 \epsilon^2)}{64 \eta^3 (\eta^2 + \epsilon^2)} - \frac{J^6 (\eta^2 + 5 \epsilon^2)}{256 \eta^5 (\eta^2 + \epsilon^2)} \notag \\
&-\frac{J^8 (25 \eta^6 - 269 \eta^4 \epsilon^2 - 405 \eta^2 \epsilon^4 - 175 \epsilon^6)}{16384 \eta^7 (\eta^2 + \epsilon^2)^3} \notag \\
&-\frac{J^{10} (49 \eta^6 + 715 \eta^4 \epsilon^2 + 1043 \eta^2 \epsilon^4 + 441 \epsilon^6)}{65536 \eta^9 (\eta^2 + \epsilon^2)^3},
\end{align}\label{d2}
respectively.

\begin{figure}[!ht]
    \centering
    \includegraphics[width=0.5\textwidth]{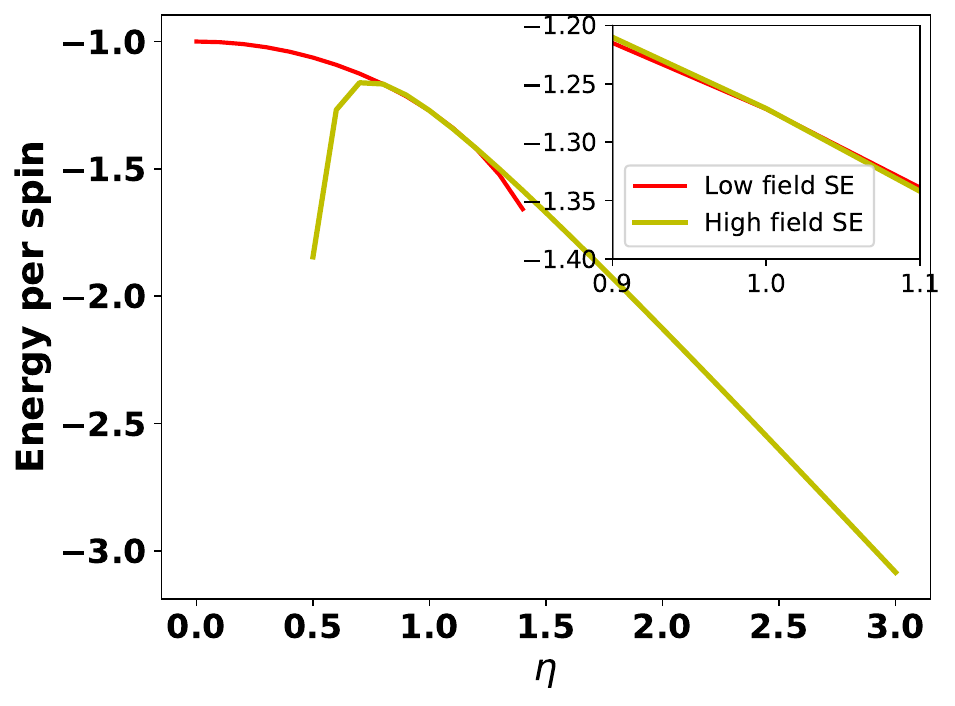}
    \caption{\textbf{SE extrapolation}. We extrapolate the SE to capture the intermediate-field regime. For both the low- and high-field regime, we remain in the region of minimal energy.}
    \label{fig12}
\end{figure}

The procedure for extrapolating the SE method in the intermediate-field regime so that these energies can be used to predict what happens involves a form of fitting. First, we choose to study the system within the range $\eta \in [0, 3]$ with $\xi = \eta/10$, where we reiterate we set $J=1$. We then subdivide this interval into two parts, where we naively define the low-field regime as $\eta \leq 1.5$ and the high-field regime as $\eta > 1.5$. Next, we plot $E_\textrm{low}^{(\eta, \xi)}$ as a function of $\eta \in [0,1.5]$ and $E_\textrm{high}^{(\eta, \xi)}$ as a function of $\eta~\in~]1.5, 3]$ with a step size of 0.1. We observe that around $\eta \approx 1.0$, the energy values of both the low- and high-field regimes begin to increase (see Fig.~\ref{fig12}). Indeed, as seen in Fig.~\ref{fig12}, for $\eta \leq 1$, $E_\textrm{low}^{(\eta, \xi)}$ remains energetically lower, i.e., below $E_\textrm{high}^{(\eta, \xi)}$, but when $\eta > 1$, $E_\textrm{low}^{(\eta, \xi)}$ becomes energetically higher. The same behavior is observed for $E_\textrm{high}^{(\eta, \xi)}$ when extrapolating backward. Assuming minimal energy in both regimes, we can infer that the low- (high-) field energy remains valid while minimizing energy up to approximately $\eta \approx 1$. Furthermore, beyond $\eta \approx 1$, $E_\textrm{low}^{(\eta, \xi)}$ ceases to exhibit the expected quadratic behavior and instead becomes linear. Similarly, for $E_\textrm{high}^{(\eta, \xi)}$, when fitted backward from $\eta = 3$, the expected linear behavior in this regime breaks down around $\eta \approx 1.0$ and transitions into a quadratic form, suggesting that the regime transition occurs at this point.

\section{Enforcing \textbf{\textit{PT}} Symmetry on the Ans\"atze}\label{A1}
In this section, we show how to enforce $PT$ symmetry in our neural network architectures. For illustration we show the implementation for an RNN. However, this implementation is valid for both MLP and RBM as well.

\begin{figure}[!ht]
    \centering
    \includegraphics[width=0.5\textwidth]{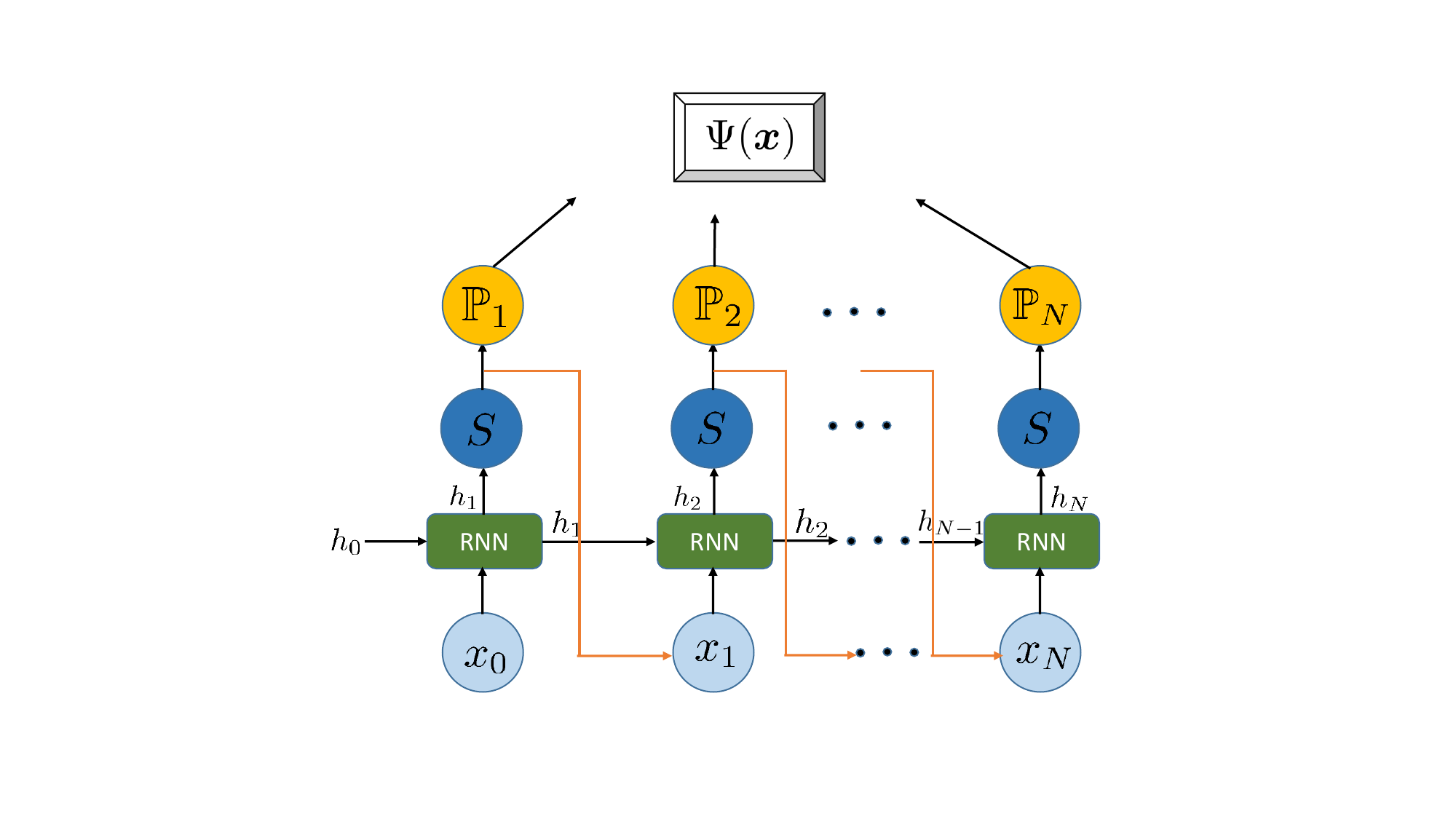}
    \caption{\textbf{RNN wavefunction}. The algorithm is fed with some spin configuration $x_i$ (light blue circles) and returns the probability distribution $\mathbb{P}_i$ (yellow circles) of that configuration, which then contributes to the final wavefunction. The green boxes represent the stacked RNN cells, and the dark blue circles are the \textit{Softmax} activation functions.}
    \label{fig11}
\end{figure}

In what follows, we propose an implementation of the RNN depicted in Fig.~\ref{fig11} that we use to find the ground-state properties of our model. We employ the approach described in Secs.~II and III in the main text as well as a method similar to the one describe in Ref.~\cite{hibat2020recurrent}. The goal is to enforce a $PT$-symmetric RNN ansatz $|\Psi_{RNN}\rangle $. Parity $P$ reflects the spatial coordinates or swaps components of a state. In a spin chain, $P$ exchange the $i^{th}$ spin with the $N-i+1^{th}$. Time reversal $T$ complex conjugates the wavefunction or parameters of the ansatz. The $PT$ symmetry is imposed on the RNN ansatz by requesting
\begin{equation}\label{EqB1}
    |\Psi_{RNN} (\bm{x}_s)\rangle = T|\Psi_{RNN} (P \bm{x}_s)\rangle,
\end{equation}
and then symmetrizing the neural network to output
\begin{equation}\label{EqB2}
    |\Psi_{RNN}(\bm{x}_s)\rangle = \frac{1}{2} \left[ F(\bm{x}_s) + F^*(P\bm{x}_s) \right],
\end{equation}
where  $F(\bm{x}_s) $ is the neural network's unregularized output and $\bm{x}_s$ the sampled spin configuration.

For a system of \textit{N} spins, let the probability distribution of the system distributed over a discrete sample space be given by $\mathbb{P}$. Let a single walker or a single snapshot of the system's states be given by a spin configuration $\bm{x}=(x_1,x_2,x_3,...,x_N)$. The probability for finding a spin in a particular configuration $\bm{x}_i$ influences the probability of finding a spin in one of the remaining configurations $\bm{x}_j = \bm{\bm{x}}-\{\bm{x}_i\}$, such that
\begin{equation}\label{EqB3}
    \mathbb{P}(\bm{\bm{x}})= \prod_{j<i=1}^N \mathbb{P}(\bm{x}_i|\bm{x}_j),
\end{equation}
where $\mathbb{P}(\bm{x}_i|\bm{x}_j)$ is the conditional probability distribution of $\bm{x}_i$ provided $\bm{x}_j$ are known. The probability in Eq.~\eqref{EqB3} is computed by the RNN cell and one can simply extract the wavefunction (see Fig.~\ref{fig11}) of the system from this probability by taking the sum of the ``square roots'' as follows
\begin{equation}\label{Eq8}
    |\Psi (\bm{\bm{x}})\rangle = \sum_{\bm{\bm{x}}} \sqrt{P(\bm{\bm{x}})}|\bm{\bm{x}}\rangle.
\end{equation}

Initially, the neural network takes as input some random spin $\bm{x}_0$ and hidden state $h_0$. Then one applies the \textit{Softmax} non-linear function to ensure that the network output has a probabilistic behavior (see Fig.~\ref{fig11}). Once the \textit{Softmax} is applied, one obtains a new spin configuration $\bm{x}_1$ and a probability amplitude or weight $\omega_1$ such that the probability at the first step is $P_1 = \omega_1 \bm{x}_1$. The new hidden state $h_1$ from this first iteration together with $\bm{x}_1$ are fed as input in the next step, and the process is repeated until the VMC recovers an energy that converges to the true ground state of the system. This is the so-called auto-regressive property of RNN. The interested reader is referred to Refs.~\cite{medsker2001recurrent, yu2019review} for more information about RNN. It is instructive to note that, provided the ubiquitous occurrence of complex-valued wavefunctions in physical systems, one can easily adapt the RNN in Fig.~\ref{fig11} to capture the complex part (phase) of the wavefunction. In this case, the RNN is designed to return both the real and imaginary parts (phase) of the wavefunction~\cite{hibat2020recurrent}. Additionally, one could train separate neural networks for the amplitude and the phase. This approach has been used in Ref.~\cite{PhysRevB.109.245120} for learning the sign structure of a frustrated Heisenberg J1-J2 model separately and has been shown to be more effective than using a neural network with complex weights.

\section{Simulation Parameters}\label{A2}

\begin{table}[!ht]
\centering
\begin{tabular}{|c|c|}
\hline
RBM parameters & Entries \\ \hline
Optimizer &  Adam \\
Seed & 111 \\
Input dimension & 2 \\
Sampling & Gibbs \\
Coupling J  & 1 \\
System size & [10,100] \\
Learning rate & $10^{-2}$ \\
Number of hidden units & 34 \\
Number of layers & 1 \\
Number of samples & 1024 \\
Training steps & $10^3$ \\
Ansatz & $PT$-sym RBM \\
Magnetic field component & $\xi = \eta/10$ \\
Offset in the log probability & $10^{-15}$ \\
Time (hh:mm:ss) for $N=10$ & $ 00:00:38$\\
Time (hh:mm:ss) for $N=10$ with TL & $ 00:00:20$\\
Time (hh:mm:ss) for $N=100$ & $ 37:00:51$\\
Time (hh:mm:ss) for $N=100$ with TL & $ 20:35:11$\\
Regularization factor of the cost function $\alpha$ & 0.1 \\
\hline
\end{tabular}
\caption{\textbf{Restricted Boltzmann Machine}. Data used for all simulations with RBM.}
\label{tab2}
\end{table}

\begin{table}[!ht]
\centering
\begin{tabular}{|c|c|} 
\hline
RNN parameters & Entries \\ \hline
Optimizer &  Adam \\
RNN cell &  Vanilla/GRU \\
Seed & 111 \\
Input dimension & 2 \\
Sampling & VMC \\
Coupling J  & 1 \\
System size & [10,100] \\
Activation function & tanh/Softmax \\
Learning rate & $10^{-2}$ \\
Number of hidden units & 34 \\
Number of layers & 1 \\
Number of samples & 1024 \\
Training steps & $10^3$ \\
Ansatz & $PT$-sym pRNN \\
Magnetic field component & $\xi = \eta/10$ \\
Offset in the log probability & $10^{-15}$ \\
Time (hh:mm:ss) for $N=10$ & $ 00:00:51$\\
Time (hh:mm:ss) for $N=10$ with TL  & $ 00:00:35$\\
Time (hh:mm:ss) for $N=100$ & $ 40:32:01$\\
Time (hh:mm:ss) for $N=100$ with TL & $ 26:57:01$\\
Regularization factor of the cost function $\alpha$ & 0.1 \\
\hline
\end{tabular}
\caption{\textbf{Recurent neural network}. Data used for all simulations with RNN.}
\label{tab1}
\end{table}

\begin{table}[!ht]
\centering
\begin{tabular}{|c|c|}
\hline
MLP parameters & Entries \\ \hline
Optimizer &  Adam \\
Seed & 111 \\
Input dimension & 2 \\
Sampling & Metropolis \\
Activation function & Relu \\
Coupling J  & 1 \\
System size & [10,100] \\
Learning rate & $10^{-2}$ \\
Number of hidden units & 34 \\
Number of layers & 1 \\
Number of samples & 1024 \\
Training steps & $10^3$ \\
Ansatz & $PT$-sym MLP \\
Magnetic field component & $\xi = \eta/10$ \\
Offset in the log probability & $10^{-15}$ \\
Time (hh:mm:ss) for $N=10$ & $ 00:00:21$\\
Time (hh:mm:ss) for $N=10$ with TL & $ 00:00:17$\\
Time (hh:mm:ss) for $N=100$ & $ 32:40:56$\\
Time (hh:mm:ss) for $N=100$ with TL & $ 20:49:00$\\
Regularization factor of the cost function $\alpha$ & 0.1 \\
\hline
\end{tabular}
\caption{\textbf{Multilayer Perceptrons}. Data used for all simulations with MLP.}
\label{tab3}
\end{table}

\section{Robustness against strong non-Hermiticity}\label{A6}

\begin{figure*}[ht!]
    \centering
    \subfloat[\centering ]{{\includegraphics[width=0.48\textwidth]{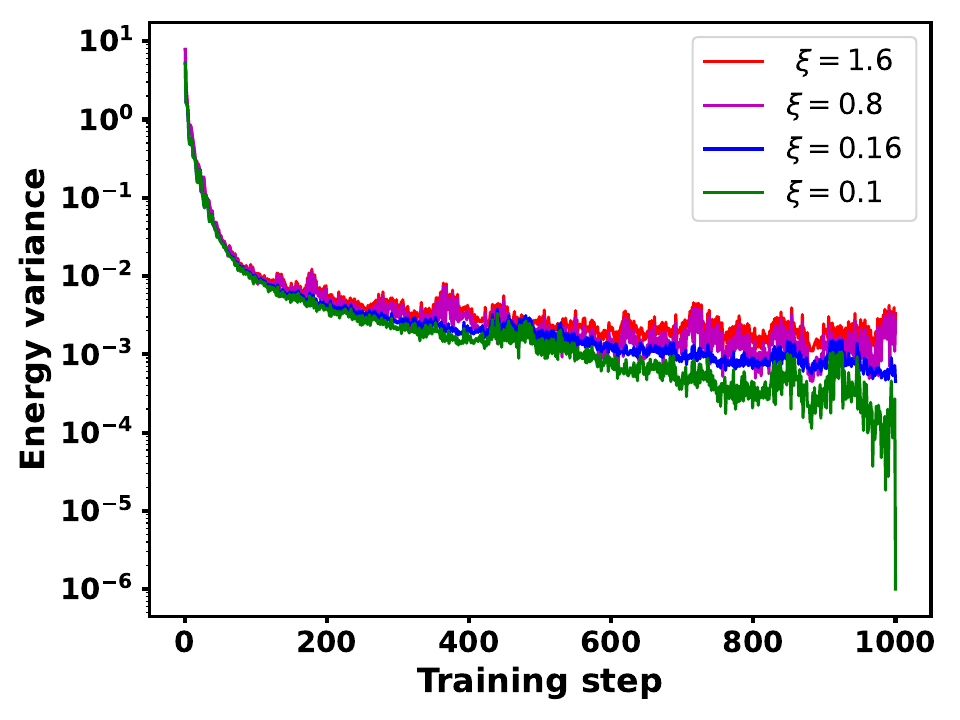} }}
    \subfloat[\centering ]{{\includegraphics[width=0.48\textwidth]{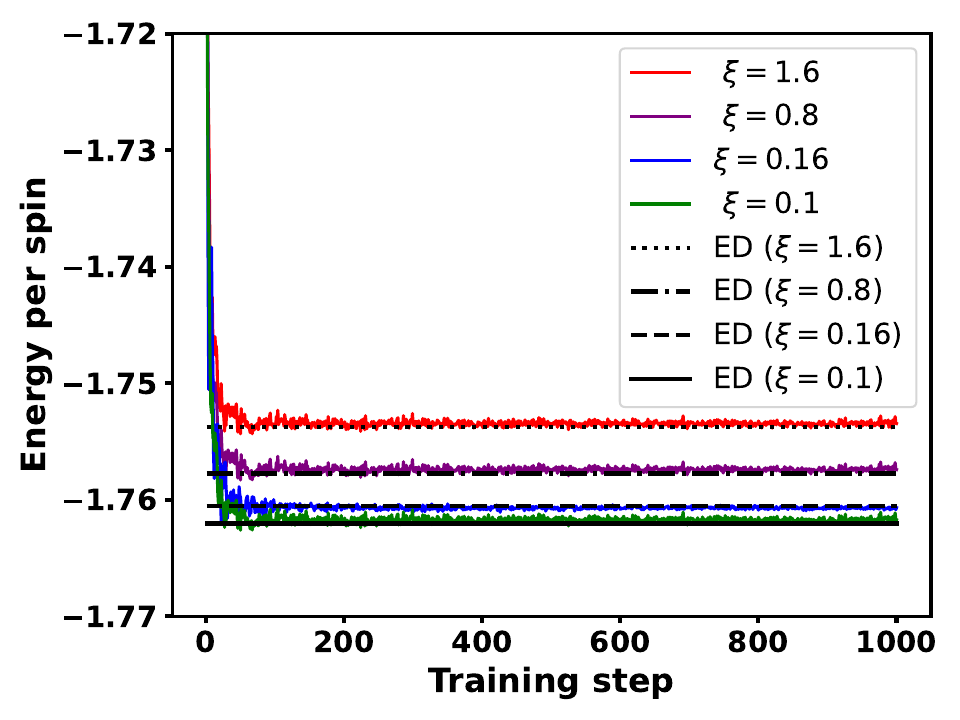} }}
    \caption{\textbf{Strong non-Hermiticity}. We plot the \textbf{(a)} energy variance, and the \textbf{(b)} energy per spin for weak and  strong non-Hermiticity. The energy variance achieved at strong non-Hermiticity (for $\eta = \xi = 1.6$) remains satisfactorily low and the algorithm converges to the ground state. $N=10$ was considered and we used the RNN. PBCs are considered in this figure.}
    \label{figs4}
\end{figure*}

It is well established that increasing the parameter governing non-Hermiticity (that is, larger $\xi$) drives the system further from the Hermitian limit. This consideration, together with our aim to remain within a physically relevant regime — where non-Hermiticity does not overwhelm the system’s behavior — motivates our focus on the weakly NH regime in this work. Nevertheless, we demonstrate in Fig.~\ref{figs4} that our NQS (RNN) remains robust even in the strongly NH case ($\eta = \xi$). In particular, we show below the energy variance achieved for $\eta = \xi = 1.6$, which remains satisfactorily low and relatively unchanged for all values of $\eta=\xi \in [0,3]$ as studied in this work. This result demonstrates that our algorithm can be reliably applied to a broad class of NH systems, effectively mitigating the instabilities typically induced by EPs. In addition, we observe that increasing the non-Hermiticity tends to shift the ground-state energy upward (increase). However, this behavior is not expected to hold universally for all NH systems, as it depends on the specific model and its symmetries.

\section{Relative errors on RNN and SE around criticality}\label{A7}

\begin{figure}
    \centering
    \includegraphics[width=0.9\linewidth]{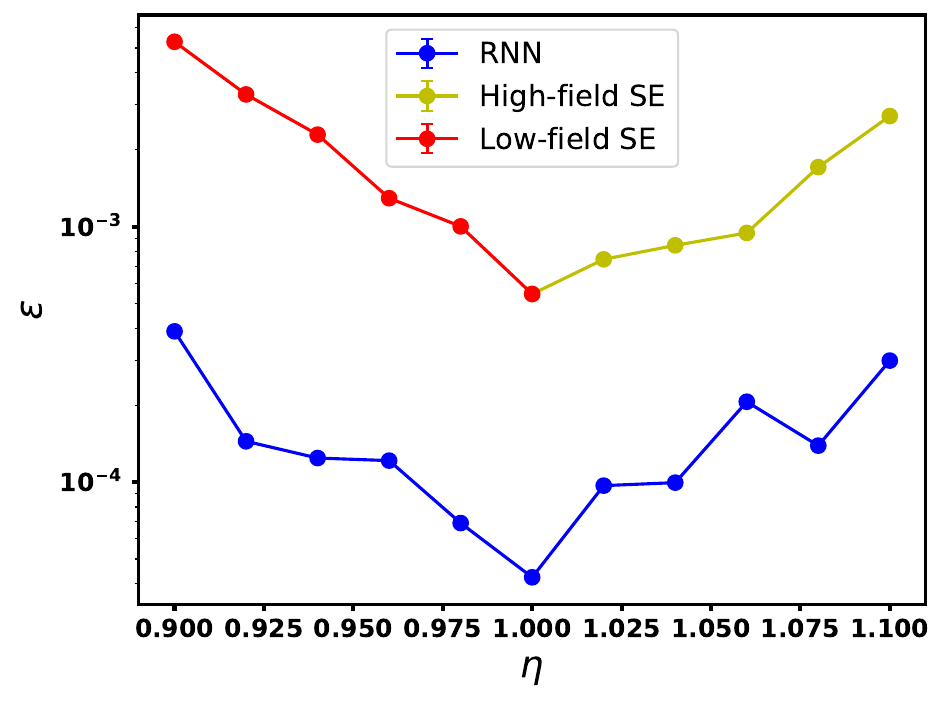}
    \caption{\textbf{Relative errors on RNN and SE around criticality}. We investigate the dip observed in the relative errors of the RNN (blue) and the SE (yellow and red). We consistently observe a dip around the transition point with both methods. We considered $N=10$ and $\xi=\eta/10$. PBCs are considered in this figure. }
    \label{figs5}
\end{figure}

Critical regions are generally expected to pose significant challenges for numerical methods, yet in our case both approaches seem to capture this point consistently. To ensure the robustness of our findings, we performed an additional analysis of our algorithm in the vicinity of $\eta = 1$, specifically examining the range $\eta \in [0.9,1.1]$, as presented in Fig.~\ref{figs5}. Within this interval, we report the observation of a dip in the relative error that our current analysis cannot fully explain. Aside from this feature, no systematic deviations or discernible patterns beyond statistical fluctuations were detected, suggesting that our approach remains stable even close to criticality. Nonetheless, given the potential importance of this regime — where NH effects may give rise to subtle spectral or dynamical signatures — a more comprehensive investigation is warranted. We regard this as an intriguing research direction.

\section{Correlation function}\label{A8}

\begin{figure*}[ht!]
    \centering
    \subfloat[\centering ]{{\includegraphics[width=0.48\textwidth]{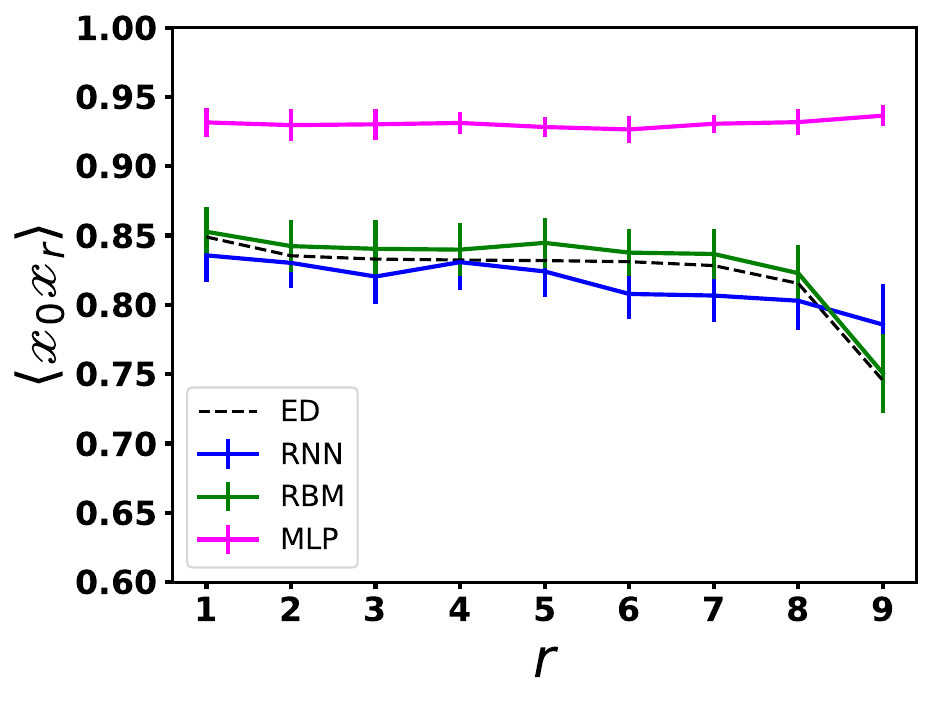} }}
    \subfloat[\centering ]{{\includegraphics[width=0.48\textwidth]{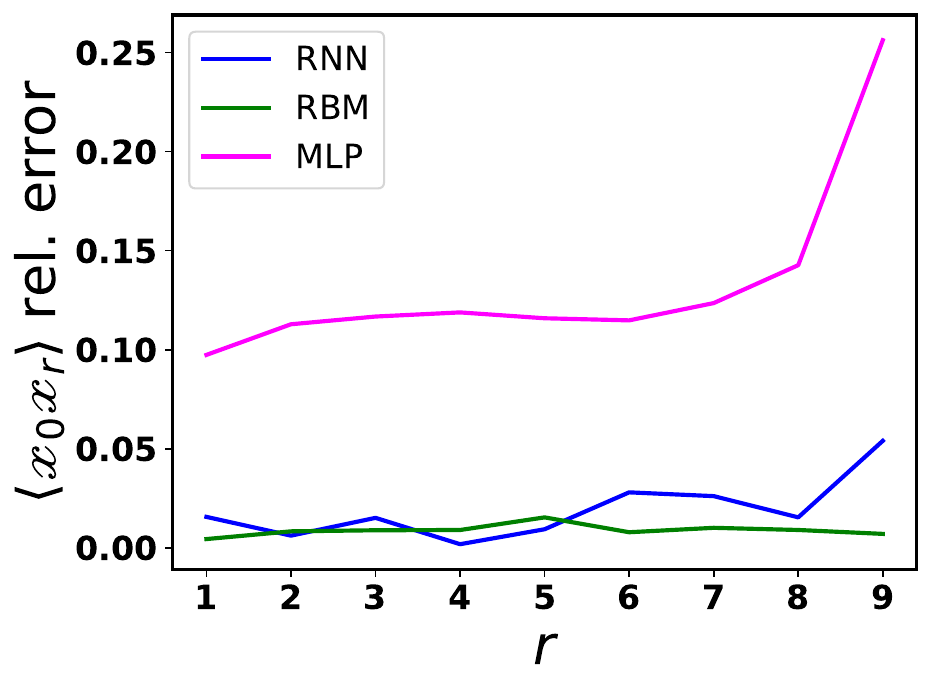} }}
    \caption{\textbf{Long-range correlations}. We plot \textbf{(a)} the long-range correlation functions using ED (black dashed line), RNN (blue), RBM(green) and MLP (magenta) and \textbf{(b)} the relative error associated to each of the NQSs. GRU is considered for RNN and $N=10$. As expected, RBM and RNN/GRU are well suited for capturing long-range correlation functions. PBCs are considered in this figure. }
    \label{figs6}
\end{figure*}

In Fig.~\ref{figs6}, we show the long-range correlation functions $\langle \bm{x}_0 \bm{x}_r \rangle$, $\forall r \in \{1,N-1\}$, for $N=10$. Vanilla RNNs are generally difficult to train for long-distance correlations due to vanishing or exploding gradients\cite{hibat2020recurrent}, a limitation alleviated by gated recurrent units (GRUs)~\cite{cho-etal-2014-properties} or long short-term memory units (LSTMs)~\cite{hochreiter1997long}. We employ a GRU RNN for our analysis; although it captures sequential correlations effectively, its accuracy for long-range correlations remains lower than that of RBMs~\cite{PhysRevB.107.195115}, which directly model the joint probability distribution and thus reproduce long-range correlations more faithfully.

\FloatBarrier

\bibliography{bibliography}

\end{document}